\documentclass[a4paper,twocolumn,11pt,accepted=2021-06-09]{quantumarticle}
\pdfoutput=1
\usepackage[utf8]{inputenc}
\usepackage[english]{babel}
\usepackage[T1]{fontenc}
\usepackage{amsmath}
\usepackage{hyperref}
\usepackage{physics}
\usepackage{tikz}
\usepackage{lipsum}
\usepackage{color}
\usepackage[numbers,sort&compress]{natbib}
\usepackage[normalem]{ulem}

\usepackage{soul}

\begin{document}

\title{Visualizing the emission of a single photon with frequency and time resolved spectroscopy}

\author{Aleksei Sharafiev}
\affiliation{Institute for Quantum Optics and Quantum Information, Austrian Academy of Sciences, Technikerstrasse 21a, 6020 Innsbruck, Austria}
\email{aleksei.sharafiev@oeaw.ac.at}
\author{Mathieu Juan}
\affiliation{Institut quantique and D\'epartement de Physique, Universit\'e de Sherbrooke, Sherbrooke, Qu\'ebec, J1K 2R1, Canada}
\author{Oscar Gargiulo}
\affiliation{Institute for Quantum Optics and Quantum Information, Austrian Academy of Sciences, Technikerstrasse 21a, 6020 Innsbruck, Austria}
\author{Maximilian Zanner}
\affiliation{Institute for Experimental Physics, University of Innsbruck, Technikerstrasse 25, 6020 Innsbruck, Austria}
\author{Stephanie W\"ogerer}
\affiliation{Institute for Experimental Physics, University of Innsbruck, Technikerstrasse 25, 6020 Innsbruck, Austria}
\author{Juan Jos\'e Garc\'ia-Ripoll}
\affiliation{Instituto de Fisica Fundamental IFF-CSIC, 28006 Madrid, Spain}
\author{Gerhard Kirchmair}
\affiliation{Institute for Quantum Optics and Quantum Information, Austrian Academy of Sciences, Technikerstrasse 21a, 6020 Innsbruck, Austria}
\affiliation{Institute for Experimental Physics, University of Innsbruck, Technikerstrasse 25, 6020 Innsbruck, Austria}

\begin{abstract}
  At the dawn of Quantum Physics, Wigner and Weisskopf obtained a full analytical description (a \textit{photon portrait}) of the emission of a single photon by a two-level system, using the basis of frequency modes \cite{weisskopf_berechnung_1930}. A direct experimental reconstruction of this portrait demands an accurate measurement of a time resolved fluorescence spectrum, with high sensitivity to the off-resonant frequencies and ultrafast dynamics describing the photon creation. In this work we demonstrate such an experimental technique in a superconducting waveguide Quantum Electrodynamics (wQED) platform, using single transmon qubit and two coupled transmon qubits as quantum emitters. In both scenarios, the photon portraits agree quantitatively with the predictions  of the input-output theory and qualitatively with Wigner-Weisskopf theory.  We believe that our technique allows not only for interesting visualization of fundamental principles, but may serve as a tool, e.g. to realize multi-dimensional spectroscopy in waveguide Quantum Electrodynamics.
\end{abstract}
\section{Introduction}
\label{intr}
The process of a photon emission in free space has been described originally by Wigner and Weisskopf \cite{weisskopf_berechnung_1930}. This simple approximation, which now can be found in nearly any quantum optics textbook, yields an expression for a joined atom-field wave function during the process of emission. In this model the atom is considered as a qubit, i.e. a two level system with ground state $\ket{g}$ and excited state $\ket{e}$, decaying in isotropic 3D vacuum. The joint wave function is given by:

\begin{equation}
\ket{\psi(t)}=e^{-t(\gamma /2-i\omega_{q})}\ket{e,0}+\ket{g}\sum_{\pmb{k}}f_{\pmb{k}}(t)\ket{1_{\pmb{k}}}
\label{eqn:WW1}
\end{equation}
\begin{equation}
f_{\pmb{k}}(t)=\frac{g_{\pmb{k}}e^{-it\omega_{\pmb{k}}}}{i\gamma/2+(\omega_{\pmb{k}}-\omega_{q})}\cdot (1-e^{-t(\gamma/2- i(\omega_{\pmb{k}}-\omega_{q}))}).
\label{eqn:WW2}
\end{equation}
The outgoing photon is shaped by the light-matter coupling $g_{\pmb{k}}$, the qubit frequency $\omega_{q}$ and the qubit spontaneous emission rate $\gamma$. After the emission process is over and the qubit is in its ground state $\ket{g}$, the field is in the state:
\begin{equation}
    \ket{f}=\sum_{\pmb{k}}f_{\pmb{k}}(t>>\frac{1}{\gamma})\ket{1_{\pmb{k}}}
    \label{f_inf}
\end{equation}
\begin{equation}
    f_{\pmb{k}}(t>>\frac{1}{\gamma})=\frac{g_{\pmb{k}}e^{-it\omega_{\pmb{k}}}}{i\gamma/2+(\omega_{\pmb{k}}-\omega_{q})}
    \label{lorentz}
\end{equation}
In the equations \ref{eqn:WW1} - \ref{lorentz} $\pmb{k}$ refers to all possible wave vectors corresponding to the modes of the electrical field, coupled to the atom. For the sake of simplicity, at this point we are using a sum in  equation \ref{eqn:WW1} instead of an integral (see appendix \ref{app_A} for detailed discussion).\par
In general, the electric field operator, considering only one polarisation direction, can be written as \cite{scully_quantum_1997}:
\begin{equation}
\label{E_gen}
\begin{split}
    &\pmb{E}(\pmb{r},t)=\pmb{E}^{(+)}(\pmb{r},t)+\pmb{E}^{(-)}(\pmb{r},t)=\\
    &=\sum_{\pmb{k}}\pmb{e_{k}}E_{\pmb{k}}a_{\pmb{k}}e^{-i(\omega_{\pmb{k}}t-\pmb{kr})}+h.c.
\end{split}
\end{equation}
where $\pmb{E}^{\pm}$ is the positive/negative frequency part of the electric field operator, $\pmb{e_{k}}$ is the unit vector along the polarization direction, $E_{\pmb{k}}$ is the mode amplitude, $a_{\pmb{k}}$ is annihilation operator in the mode with wave vector ${\pmb{k}}$ and $\pmb{r}$ is the observation point location. State from equation \ref{f_inf} has zero expectation value for the field. In superposition with the vacuum state, the electric field can be calculated as 
\begin{equation}
    <\pmb{E}>\propto\bra{0}\pmb{E}^{(+)}\ket{f}\propto\sum_{\pmb{k}}\sqrt{\hbar\omega_{\pmb{k}}}f_{\pmb{k}}(t)e^{i\pmb{kr}}.
    \label{E}
\end{equation}
Following \cite{scully_quantum_1997} and making the usual assumption that the frequencies of interest are located around $\omega_{q}$, one can exchange  $\sqrt{\omega_{\pmb{k}}}$ with $\sqrt{\omega_{q}}$ and take it out of the sum. In  this approximation, it turns out that $f_{\pmb{k}}(\omega_{\pmb{k}},t)$ defines the time-dependent wavepacket, i.e field amplitudes for a certain $\omega_{\pmb{k}}$. After exchanging the sum in expression \ref{E} with an integral, one arrives at a usual truncated exponential decay.

\begin{figure}[!t]
    \centering
    \includegraphics[width=1.0\linewidth]{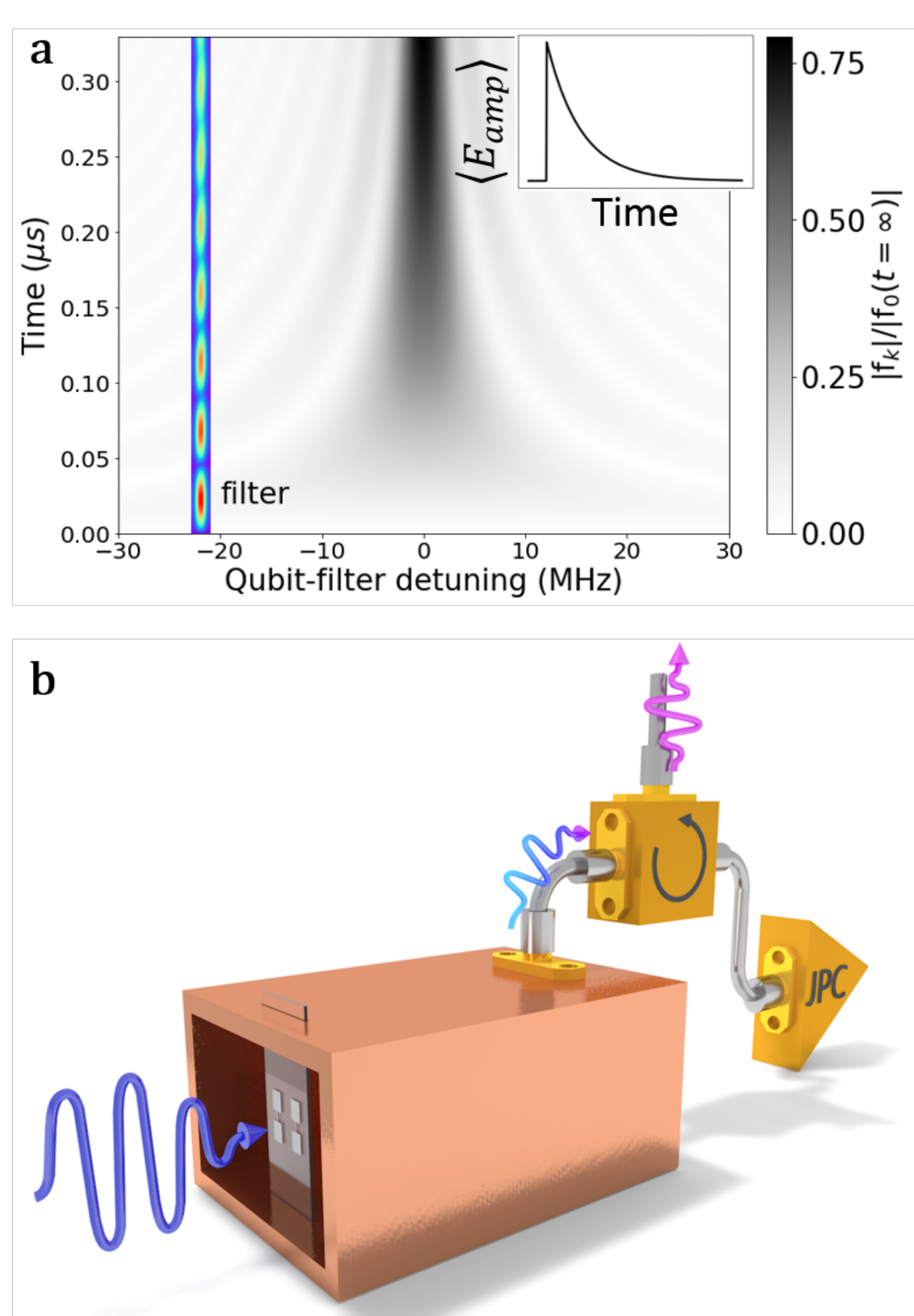}
    \caption{The idea of the visualization of the photon emission with frequency and time resolved spectroscopy. \textbf{a}, Visualization of formula \ref{eqn:WW2}: single photon spectrum, as a function of time, i.e. set of $|f_{\pmb{k}}(t)|$ in Wigner-Weisskopf theory, which is proportional to the electric field amplitude $E_{\pmb{k}}$, corresponding to the wave vector $\pmb{k}$. Along the x-axis we plot the qubit detuning from the mode $\pmb{k}$: $\omega_{\pmb{k}}-\omega_{q}$. The plot is normalized by $|f_{0}(t=\infty)|$, corresponding to zero detuning. For the plot we used $\gamma/2\pi=1.5$MHz and assumed g$_{\pmb{k}}$ to be constant within the vicinity of the qubit frequency. The colored area is strongly amplified by the narrow-band quantum limited amplifier (JPC), playing the role of the filter. The JPC allows to visualize the entire picture line by line by sweeping the qubit detuning relative to the JPC resonance. The inset depicts the electric field amplitude $<E_{amp}>$ (integrated over all frequencies) of the same signal. \textbf{b}, Experimental setup. Two transmon qubits are coupled to the fundamental mode of a copper waveguide with rectangular cross section. The coherent drive is applied to the qubits through the waveguide (blue wavy arrow). After the abrupt stop of the drive the system emits a relatively broadband signal (see the inset on the subplot \textbf{a}). The particular frequency component which is being amplified by the JPC is shown by the pink wavy arrow.}
    \label{fig:exp_idea}
\end{figure}

Figure \ref{fig:exp_idea}(a) shows $|f_{\pmb{k}}|$, which is proportional to electric field amplitude, as a function of both qubit-filter detuning $\omega_q-\omega_{\pmb{k}}$ and time $t$. At the very first instant of the emission process the field state is distributed over a wide frequency range, exemplifying Heisenberg uncertainty principle. After sufficient time ($>\frac{1}{\gamma}$) it narrows down to a usual Lorentz-shaped line, described by equation \ref{lorentz}. Inset in this picture shows the same wavepacket after integrating over all $\omega_{\pmb{k}}$ (the truncated exponent). The initial spectral broadening is a result of the sharp rise of the field in the inset. The spectral components of the wavepacket from the main plot can be calculated performing a Fourier transform of the field shown in the inset, taken in a time window from $-\infty$ to $t$. At long times $t\rightarrow \infty$ the spectrum corresponds to a single Lorentz-shaped peak. In this paper we demonstrate a circuit Quantum Electrodynamics technique which allows to experimentally measure \{$f_{\pmb{k}}$\} directly without any post processing.

Before describing our approach it is important to discuss shortly the physical meaning of the expression \ref{E}. As it describes field amplitudes of a single photon signal it is sometimes associated with a \textit{quantum wave function of a photon} $\Psi(\pmb{r},t)$ (as mentioned in \cite{scully_quantum_1997} and discussed in more details in \cite{sipe_photon_1995,bialynicki-birula_photon_1994,bialynicki-birula_photon_1996}). In this case $|\Psi(\pmb{r},t)|^2$ has an intuitively clear meaning of the energy density of the photon. This or similar terminology is sometimes used in literature, including experimental papers \cite{lundeen_direct_2011,davis_measuring_2018}. This may stir some controversy related to localization problems of massless spin 1 particles first pointed out by Newton and Wigner \cite{newton_localized_1949}. Interestingly though, this problem only appears for a spatial wave function, while in momentum (frequency) space the single photon wave function is a well defined mathematical object. Nevertheless, to avoid any terminology related confusions, in this paper we avoid using "single photon wave function", and refer to our experimental result, i.e. reconstruction of $f_{\pmb{k}}(\omega_{\pmb{k}},t)$ through multi-modal measurements of electric field, as "single photon portrait".

To reconstruct the portrait experimentally we use a waveguide with a rectangular cross section. As discussed in Appendix \ref{app_A}, the Wigner-Weisskopf approach still holds in this quasi-1D environment, providing that the qubit frequency is not too close to the cutoff frequency of the waveguide. Therefore, our approach is well described by the theory of microwave photons propagating through quasi-one-dimensional waveguides, with no polarization or band degree of freedom. In general, regardless whether the Wigner-Weisskopf approximation is applicable, a single photon state $\ket{F}$ can be written in momentum space as a linear combination of single-photon eigenstates
\begin{equation}
  \ket{F} = \sum_{\pmb{k}} F_{\pmb{k}}(t)\ket{1_{\pmb{k}}}= \sum_{\pmb{k}} F_{\pmb{k}}(t)\,a_{\pmb{k}}^\dagger\ket{0}.
  \label{f_k}
\end{equation}
The photons are labeled by momenta $\pmb{k}$ and we assume again that there are no other degrees of freedom. The coefficients $F_{\pmb{k}}$ can be reconstructed as $F_{\pmb{k}}=\mel{0}{a_{\pmb{k}}}{F}$, and coincide with $f_{\pmb{k}}(t)$ from \ref{eqn:WW2} when the Wigner-Weisskopf model is valid. Considering photons propagating only in one direction in the waveguide, wave vector $\pmb{k}$ can be substituted by wave number $k$.  Our protocol to reconstruct the photon portrait (i.e. reconstruct $F_{\pmb{k}}(t)=F_{k}(t)$) consists of 3 steps: 1) create a state with a slight superposition of vacuum and a single photon state, $\ket{\Psi}=C_{1}\ket{0}+C_{2}\ket{F}$; 2) pass this through a filter that selects a narrow band of momenta around $k_{filter}$ (and corresponding frequency $\omega_{filter}$) creating the state $\rho = P\dyad{\Psi}{\Psi}P+[1-\mel{\Psi}{P}{\Psi}]\ketbra{0}{0}$, where $P$ is the projection operator on a narrow bandwidth $\delta k$ ($\delta\omega$) around $k_{filter}$; 3) measure the field $E(t)\propto F_{k_{filter}}(t)$. In the paper we refer to this reconstruction as \textit{photon portrait} visualization. Figure \ref{fig:exp_idea}(a) illustrates this procedure when applied to a usual Markovian situation, where the Wigner-Weisskopf approximation is valid: after passing a narrow band filter, highlighted by the colored area electric field $E(\omega_{filter},t)$ is being recorded and averaged. The process is repeated for different detunings between the atom ($\omega_{q}$) and the filter ($\omega_{filter}$). Note, that the procedure itself by no means is limited to the standard Wigner-Weisskopf emission picture with fixed qubit-environment coupling rate $\gamma$ and corresponding spectrum from the figure \ref{fig:exp_idea}(a). In fact, in this simplest form this approximation often does not hold in typical experimental situations. For instance, the photon envelope from the inset on the figure \ref{fig:exp_idea}(a) can be artificially controlled, for example to achieve symmetrically shaped photons for quantum information transfer \cite{srinivasan_time-reversal_2014,pechal_microwave-controlled_2014}. Our protocol to reconstruct $F_{\pmb{k}}(t)$ is still applicable in this case.  

\section{\label{res}Results}
In the following we describe how $F_{\pmb{k}}(t)$ can be experimentally measured with a state-of-art superconducting waveguide Quantum Electrodynamics (wQED) platform, using multi-mode frequency and time-domain resolved spectroscopy.

More generally, superconducting circuit QED is a particularly promising platform for experiments in quantum optics and quantum physics generally due to its flexibility and readily available set of microwave tools. Possible applications include exploring the ultra/deep-strong coupling regimes of light-matter interaction \cite{forn-diaz_ultrastrong_2016, langford_experimentally_2017, braumuller_analog_2017}, interaction with non-Ohmic environments \cite{liu_quantum_2017,bronn_broadband_2015,hoi_probing_2015} and cooperative effects \cite{mlynek_observation_2014, mirhosseini_cavity_2019}, see also review \cite{wendin_quantum_2017}.
Furthermore, by using the superconducting circuit QED platform we are taking advantage of quantum limited parametric amplifiers. These, or similar devices, have become a standard part of any experimental circuit QED setup. In particular they add very little noise to the amplified signal: reduced to the input noise can be as low as half of a photon in terms of added noise number \cite{caves_quantum_1982}. In our setup we are using an amplifier known as Josephson Parametric Converter (JPC)\cite{bergeal_phase-preserving_2010}. The limited amplification bandwidth of a JPC ($\sim$5\,MHz), which in many cases can be considered as a drawback, allows to effectively Fourier filter the single photon signal.

The central part of our experimental setup consists of two transmon qubits placed into a copper rectangular waveguide with a fundamental mode cutoff frequency at $\sim$\,6.5\,GHz and a second mode cutoff at $\sim$\,13\,GHz (shown in figure \ref{fig:exp_idea}(b)). The transmons are installed in the same plane relative to the phase front of the fundamental mode of the waveguide. The waveguide is thermally anchored to the base plate of a dilution refrigerator with temperature $\sim$\,20\,mK. One end of the waveguide is used as input and is connected to a heavily attenuated coaxial line coming from room temperature electronics. The other end of the waveguide is connected to the JPC and consequently to an output line, featuring a conventional transistor-based amplification chain (see Appendices \ref{app_B} and \ref{app_C} for the details). Our emitter consists of two frequency-tunable (between 6.1 and 8.5\,GHz) transmon qubits installed inside the waveguide and thermalized via a copper clamps. The qubit frequencies are independently controlled by two superconducting coils installed on the external surface of the waveguide (not shown). The transmons were designed to be identical, with a total length of 1.7\,mm each and a nonlinearity of $\sim$\,220\,MHz. One of the key features of circuit QED platform is that qubit-qubit and qubit-environment couplings can be designed independently. Indeed, transmon qubits in this architecture interact with each other as electric dipoles \cite{dalmonte_realizing_2015} and therefore the coupling depends on their mutual position and orientation. The qubit-environment interaction, in turn, within the same electric dipole approximation depends on the qubit position in the waveguide and its orientation with respect to the electric field vector of the fundamental mode of the waveguide. Exact values for both interactions depend on geometry and can be obtained from Maxwell equations with any finite element solver. In our setup the qubits are installed in parallel to each other and symmetrically in the center of the waveguide providing equal coupling to the waveguide for both transmons. Using the possibility of independent engineering of qubit-qubit and qubit-waveguide couplings in our setup, we have designed the direct coupling between the qubits to be $\sim$60\,MHz, exceeding the coupling between the qubits and the waveguide $\frac{\gamma}{2\pi}\sim$1.5\,MHz. This is necessary to observe collective effects in our system, as it gives the qubits time to interact before the system decays radiatively into the waveguide.

The control signal quadratures were independently generated by two channels of an Arbitrary Waveform Generator with 1\,GS/s sampling rate at a frequency of 200\,MHz and consequently upmixed with a conventional IQ-mixer to the qubit frequency. The time-domain signal at the qubit frequency was downmixed (after JPC-filtering and amplification) to an intermediate frequency of 50\,MHz before being recorded in time domain with a 16 bit resolution and 800\,Msamples/s rate Digital-to-Analog converter. In a first set of experiments, we switch the JPC pump off, and do some preliminary measurements without filtering on a single qubit.
\begin{figure}[t]
\centering
\includegraphics[width=1.0\linewidth]{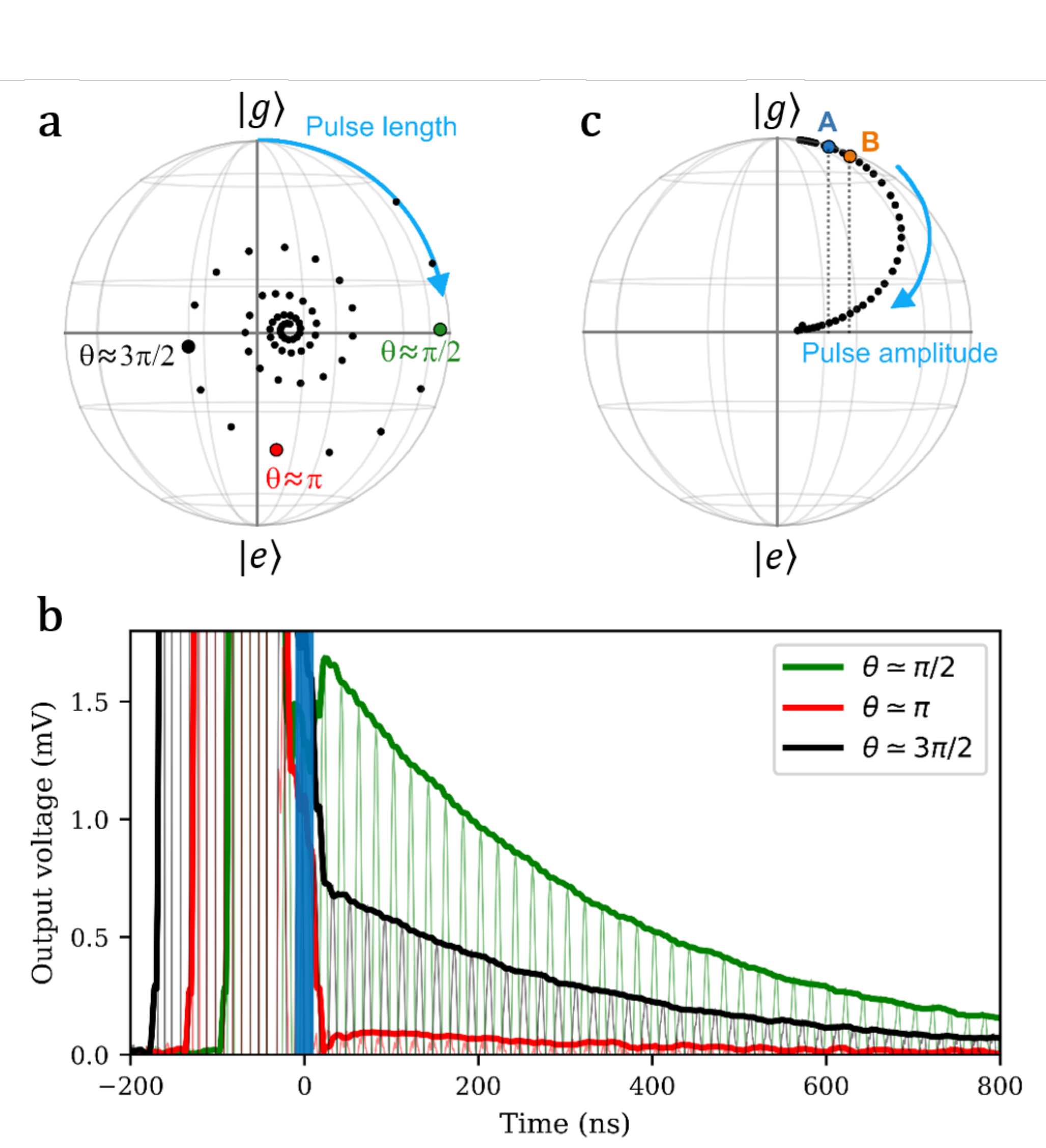}
\caption{Qubit dynamics under short and long resonant drive. \textbf{a}, Reconstructed qubit state (in the rotating frame) on the Bloch sphere as a function of drive length. \textbf{b}, Averaged measured electric field $\langle a_{out}(t) \rangle$ (which is $\sim\langle \sigma^{-}(t) \rangle$ after the pulse stop) detected at the output of the waveguide, while driving the qubit resonantly to 3 different rotation angles $\theta$ marked in subplot (a). Thick lines are the extracted envelopes from the raw data (thin lines). The pulses are shifted in time to stop at the same time moment moment $t=$\,0\,ns. For better contrast only positive voltages are shown. The data was taken without filtering, i.e without JPC. After the drive stop these data essentially correspond to the inset in the figure \ref{fig:exp_idea}(a) \textbf{c}, Steady state curve inside the Bloch sphere where the qubit state ends up after "infinitely long" pulse. The marked points A and B correspond to different ratios of the drive strength $\alpha$ to the square root of the qubit radiative decay rate $\sqrt{\gamma}$: $[\frac{\alpha}{\sqrt{\gamma}}]_{B}/[\frac{\alpha}{\sqrt{\gamma}}]_{A}\approx\sqrt{2}$}.
\label{fig:bloch_spheres}
\end{figure}

Passing through the waveguide, the input signal $\alpha e^{-i\omega t}$ interacts resonantly with the qubit inside, changing both its population and coherence. Due to the presence of the drive and qubit internal losses, which are not included in the Wigner-Weisskopf theory, the system is better described using the input-output formalism \cite{lalumiere_input-output_2013}. Being an open quantum system the qubit state in the rotating frame follows a spiral trajectory inside the Bloch sphere during the pulse, as shown in figure \ref{fig:bloch_spheres}(a). Recording the output signal $\langle a_{out}(t) \rangle$ after the drive is abruptly stopped and fitting it with an input-output theoretical model (see Appendix \ref{app_E}) allows to effectively measure the qubit coherence $\langle \sigma^{-} \rangle$ as a function of time. Having access to $\langle \sigma^{-}(t) \rangle$ during the drive pulse as well as the decay and assuming that the qubit is initially in the ground state $\ket{g}$, one can access the qubit population $\langle \sigma^{z}(t) \rangle$ as well.
This information is enough for a qubit state tomography on an open quantum system. Similar protocols have already been implemented in circuit QED systems \cite{houck_generating_2007,abdumalikov_dynamics_2011,eichler_experimental_2011}.

Figure \ref{fig:bloch_spheres}(b) shows the raw detected signal (still in the absence of JPC-filtering) from the qubit for three pulses of different length, approximately corresponding to rotating the qubit state by $\pi/2$, $\pi$ and $3\pi/2$. The qubit state after each of these driving pulses correspond to the large green, red and black dots in the Bloch sphere shown in figure \ref{fig:bloch_spheres}(a) respectively. Right after the drive pulse, the envelope of the detected signal (thick lines in figure \ref{fig:bloch_spheres}(b)) is proportional to $\left<\sigma^{-}\right>$. 
As a result, the signal from the qubit decay is largest for a $\theta \approx \pi/2$ drive pulse which maximises the qubit coherence at the end of the pulse; conversely the signal is almost non-existent for $\theta \approx \pi$. After the drive stop the envelopes of these signals essentially correspond to the inset of the figure \ref{fig:exp_idea}(a).
\begin{figure}[!b]
\centering
\includegraphics[width=1.0\linewidth]{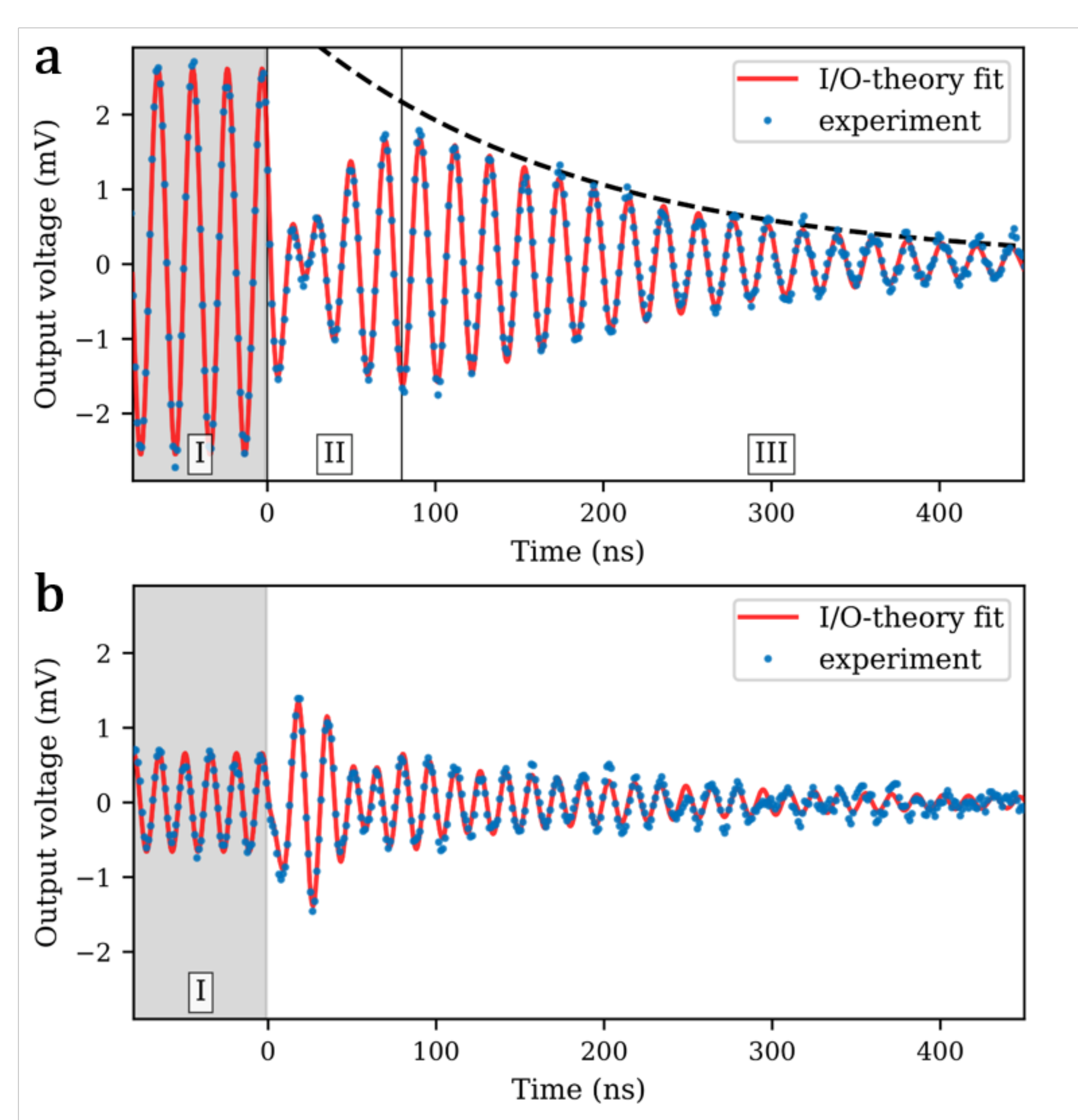}
\caption{JPC-filtered qubit dynamics. \textbf{a}, Raw down-mixed experimental signal for zero JPC-qubit detuning after an "infinitely long" drive pulse (blue dots) and the theoretical fit for it based on input-output theory (red line). The signal visually can be separated in 3 zones:  zone I (negative time) corresponds to the last portion of the drive, zone III corresponds to the qubit exponential decay (dashed line) amplified by the JPC. Zone II is a result of a destructive interference between the photon, stored in the qubit and the photons stored in the JPC at the moment of the drive stop. \textbf{b}, same as \textbf{a} for a $17$~MHz JPC-qubit detuning.}
\label{fig:single_qubit_trace}
\end{figure}

For the particular case of an "infinitely long'' pulse (much longer than $\gamma^{-1}$) the qubit reaches dynamic equilibrium during the pulse and the data can be fit using an analytical input-output model (see Appendix \ref{app_D}). Experimentally we used pulses with a length of 1$\mu s>\gamma^{-1}\approx (2\pi\cdot1.5)^{-1}\mu s$. After such a pulse, the possible qubit states are restricted to an "arc" inside the Bloch sphere (see figure \ref{fig:bloch_spheres}(c)). Its exact position along the "arc" depends on the ratio of the drive amplitude $\alpha$ to the square root of the qubit radiative decay $\sqrt{\gamma}$. 

In order to visualize the emission process we need to reconstruct all $f_{\pmb{k}}(t)$. To do this we switch on the JPC pump, and use the JPC as a filter centred at $\omega_{\pmb{k}}$. Therefore we have access to $\left<\sigma^{-}\right>$ over a narrow frequency band effectively measuring one $f_{\pmb{k}}(t)$: sweeping the JPC-qubit detuning allows us to fully reconstruct the photon portrait. For the case of zero detuning, figure \ref{fig:single_qubit_trace}(a) shows a JPC-filtered signal where the qubit was excited using an "infinitely long" pulse. After the pulse is switched off (at the end of zone I), the JPC and the qubit fields are interfering destructively resulting in a minimum in the signal at $\sim$30\,ns followed by a revival (zone II) and finally an exponential decay of the qubit (zone III). The constriction in zone II comes from the finite JPC memory time, given by its bandwidth. At the moment of the drive stop $t$\,=\,0~ns both the qubit and the JPC store some energy. With the chosen - rather strong - drive amplitude, JPC is mostly populated with the photons which passed the qubit without interacting with it (avoiding a phase shift associated with the qubit absorption and reemission). As the JPC decays faster than the qubit, this affects only the signal within the zone II. It is not captured directly by Wigner-Weisskopf as this model assumes empty electromagnetic modes before the emission. Conversely, this effect is naturally appearing within the input-output approach. For the case where the JPC is detuned from the qubit, figure \ref{fig:single_qubit_trace}(b), the signal does not show the simple exponential decay during the qubit emission, so we do not emphasize regions II and III in this figure. Independently of the detuning, since the qubit is prepared using an ''infinitely long'' pulse we fit the data in the whole range with the analytical input-output model (red-lines in figure \ref{fig:single_qubit_trace}, see Appendix \ref{app_D} for the model), reproducing accurately all the features in the signals.
\begin{figure*}[t]
\centering
\includegraphics[width=\textwidth]{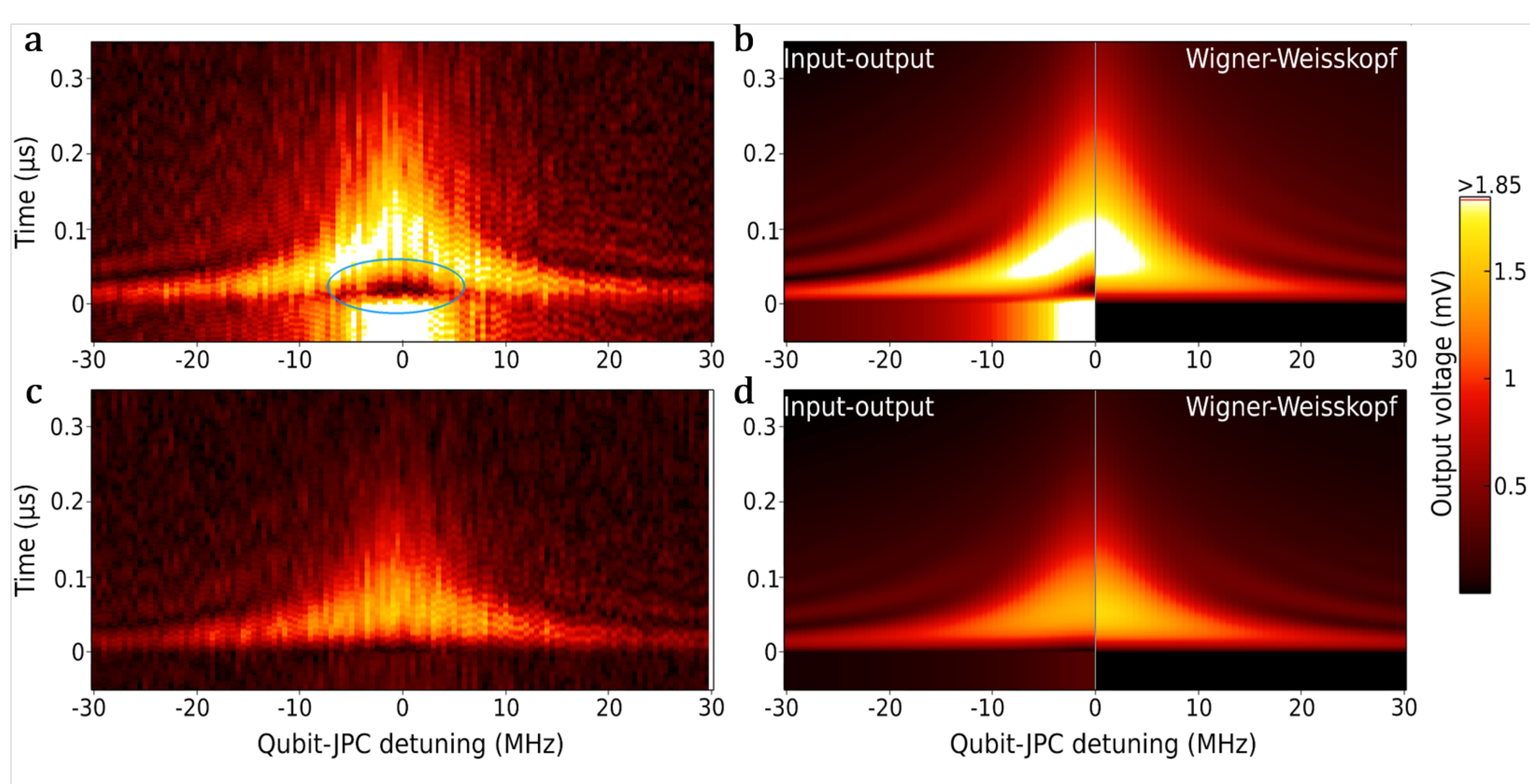}
\caption{Single photon portraits obtained with frequency and time domain resolved spectroscopy. \textbf{a}, \textbf{b}: experimental and theoretical portraits of a single photon emitted by the transmon qubit into the waveguide. The two sides of \textbf{b} correspond to the input-output approach (left) and pure Wigner-Weisskopf picture (right). The envelop of the signal shown in figure \ref{fig:single_qubit_trace}(a) is a cut of subplot (a) of this figure corresponding to zero qubit-JPC detuning  \textbf{c},\textbf{d}, photon emitted from a collective bright state of two directly coupled transmons. The experimental result \textbf{c} is closer to the Wigner-Weisskopf picture, than in the case of a single transmon.  }
\label{fig:wf}
\end{figure*}
Sweeping the JPC-qubit detuning allows us to reconstruct the \textit{photon portrait}: the frequency resolved time evolution of the electric field amplitude, shown in figure \ref{fig:wf}(a), where the color code corresponds to the envelope of the detected signal. The corresponding theoretical pictures are shown in figure \ref{fig:wf}(b). The left side of this plot was obtained with the input-output approach, which matches the experiment well, and the right side is the pure Wigner-Weisskopf picture. The destructive interference due to the JPC memory is visible in the region highlighted in figure \ref{fig:wf} with a blue oval. To obtain the portraits closer to the pure Wigner-Weisskopf picture one has to ensure that all photons in the detected signal interacted with the qubit. There are two possibilities to do it: one can either decrease the drive strength $\alpha$ to give the qubit more time to relax into the waveguide before it has to "process" the next photon, or increase $\gamma$ and therefore speeding up the "processing" rate. Both ways are relatively easy to realize experimentally, nevertheless we choose the second approach for the following reasons: First, increasing $\gamma$ also decreases the influence of non-radiative losses on the dynamics. Second, decreasing $\alpha$ would also decrease the detected signal amplitude, which would in turn require us to increase the number of averages to achieve the same signal-to-noise ratio. Finally, we believe that it is relevant to demonstrate how the method of frequency and time resolved spectroscopy works in the case of more complex systems than single qubits, demonstrating that this approach is useful to study open quantum systems.  

In order to control the qubit-environment coupling $\gamma$ in-situ, we are taking advantage of the second qubit inside the waveguide, tuning it into resonance with the first. Being placed in the waveguide symmetrically, the system forms one bright and one dark state (see Appendix \ref{app_F}). The bright state couples to the environment twice as strongly as a single qubit. The protocol described for the single qubit case above, was implemented as well on the bright state of the two-qubit system. Figure \ref{fig:wf}(c) shows the experimental result, while the corresponding theoretical pictures (input-output and Wigner-Weisskopf) are shown on the two sides of figure \ref{fig:wf}(d). Clearly, in this case the experimental result is closer to the Wigner-Weisskopf case, than in figure \ref{fig:wf}(a). In the area with the drive still on ($t<0$), the signal now is almost absent because the bright state forms a better mirror than a single qubit, due to the increased ratio of the radiative decay to the internal losses. In addition, as the drive strength $\alpha$ was kept the same in both cases (figures \ref{fig:wf}(a) and (c)), the difference in $\gamma$ effectively changed the steady state population and excited the effective "qubit" to different points inside the Bloch sphere, illustrated with points A (two-qubits superradiant state) and B (single qubit) in the figure \ref{fig:bloch_spheres}(c). This leads to an effectively weaker drive for the collective case. We note that the noise visible in the experimental data is in part due to fluctuations in the qubit frequency. This fluctuations lead to unwanted qubit-drive detuning, which for the theory pictures \ref{fig:wf} (b) and (d) was assumed to be zero. This detuning, though, can be taken into account within the same input-output model, see Appendix \ref{app_D} for the details and corresponding pictures. The rest of the noise, especially at long times, in the figures \ref{fig:wf}(a) and (c) could be in principle removed by longer averaging. This noise is especially noticeable at large detunings, as we have to detect weak signals. For this setup number of averages (10$^5$ times) represents a compromise between the averaging time, during which we assume the qubits to be perfectly frequency stable, and the noise level.

One clear difference between the theoretical predictions of {$f_{\pmb{k}}(t)$} from figure \ref{fig:exp_idea}(a) and the experimental portraits from figure \ref{fig:wf} is that the measured signal dies out after the qubit has completely relaxed. This is related to the fact that we are measuring the signal at one fixed position of the detector and JPC is an imperfect Fourier filter with finite linewidth. Accordingly, the "Wigner-Weisskopf portraits" from the right halves of the figures \ref{fig:wf}(b) and (d) were obtained by inserting the time domain signal  from the inset of the figure \ref{fig:exp_idea}(a) into the finite bandwidth filter function, defined by the equation \ref{filter_function} from the Appendix \ref{app_D}. 
\section{\label{con}Conclusion}
In this work we have developed an experimental method to implement frequency and time resolved spectroscopy. We have demonstrated the power of this technique by obtaining a time-resolved portrait of a very fundamental process: the emission of single photons. All the essential features of the Wigner-Weisskopf picture - narrowing down of the spectrum, time-energy uncertainty relation - are captured by the experiment, which illustrates the deep physical connection between the Fourier transform and Heisenberg's uncertainty principle.

In comparison with some other experiments reported earlier (e.g. \cite{houck_generating_2007}, \cite{eichler_experimental_2011}, \cite{mallet_quantum_2011}) we concentrate on the spectral components of a single photon signal, especially at the first stage of the emission process. This allows in particular to probe the qubit environment around the frequency of the qubit. A "usual" Wigner-Weisskopf picture observed experimentally in this paper, is a base line for this technique. Deviation from this picture would signify more complex physics involved.  

In this context, we anticipate the method of frequency and time resolved spectroscopy to become a convenient tool to characterize circuit-QED systems and especially open quantum systems like waveguide QED. It can be used to study not just single emitters, but Markovian and non-Markovian effects in multi-level emitters and many-body quantum simulators. Furthermore, by splitting the output port, the same setup can be upgraded to perform complete tomography of multi-photon scattering \cite{ramos2017}.

This method offers, for instance, a possibility to reveal and perhaps identify a two-level system (TLS) weakly  coupled to a qubit - a crucial problem in the context of circuit QED - even if there is no possibility to tune the qubit or TLS frequencies and observe the TLS with conventional spectroscopy. TLS with frequencies not exactly resonant with the qubit, but close to it, might be harmful for the device performance. Even if there is a possibility to record the qubit decay in time domain, it might be difficult to reveal the presence of an off-resonant TLS. As only one particular frequency is being amplified by the JPC, the weak off-resonant features, which would be otherwise covered by noise or signal at the central frequency, can be resolved after relatively modest averaging. In fact, this method can be compared to some of the known 2D fluorescence techniques in optics, e.g. laser-induced fluorescence \cite{chen_introduction_2016}. This similarity points out that various types of multi-dimensional spectroscopy methods can be straightforwardly developed for circuit QED systems on the basis of our frequency-time spectroscopy.

\section{Acknowledgements}
\label{ackn}
AS and MZ are funded by the European Research Council (ERC) under the European Unions Horizon\,2020 research and innovation program (grant agreement No 714235). MZ is also supported by the Austrian Science Fund FWF within the DK-ALM (W1259-N27). MLJ is funded by MaQsens project (European Union’s Horizon\,2020 research and innovation program, grant agreement No 736943 and European Research Council under the European Union’s Horizon\,2020 research and innovation program, grant agreement No 649008). J.J.G.R. acknowledges support from project PGC2018-094792-B-I00 (MCIU/AEI/FEDER, UE), CSIC Research Platform PTI-001, and CAM/FEDER project No. S2018/TCS-4342 (QUITEMAD-CM). Facilities use was supported by the KIT Nanostructure Service Laboratory (NSL).
\bibliographystyle{unsrtnat}
\bibliography{quantum}
\onecolumn\newpage
\appendix
\textbf{\Large APPENDICES}
\section{Wigner-Weisskopf model in  a rectangular cross section waveguide}
\label{app_A}
In this section we discuss to what extend the Wigner-Weisskopf approach, developed originally for an atomic decay in 3D vacuum is relevant to our experimental situation of a rectangular cross section waveguide.

We follow \cite{scully_quantum_1997} and start from the very general equation for the state vector of the atom-photon system:
\begin{equation}
    \ket{\Psi(t)}=c_{e}(t)\ket{e,0}+\sum_{\pmb{k}}c_{g,\pmb{k}}(t)\ket{g,1_{\pmb{k}}}
\end{equation}
with initial condition $c_{e}(0)=1$ and $c_{g,\pmb{k}}(0)=0$. We are going to use the standard Hamiltonian in the interaction picture and within rotating wave approximation:
\begin{equation}
    H=\hbar\sum_{\pmb{k}}(g_{\pmb{k}}^{*}\sigma_{+}a_{\pmb{k}}e^{i(\omega_{q}-\omega_{\pmb{k}})t}+h.c.)
\end{equation}
where $a_{\pmb{k}}$ is the annihilation operator for the field mode with wave vector $\pmb{k}$, $g_{\pmb{k}}=g_{\pmb{k}}(\pmb{z_{0}})=g_{\pmb{k}}(0)e^{-i\pmb{k}\pmb{z_{0}}}$ is the coupling between the field mode $\pmb{k}$ and the atom positioned in the waveguide at the point $z_{0}$,  and h.c. refers to hermitian conjugate. In the further discussion we will assume that the atom is located at the point $z_{0}=0$. The time dependent Schr\"odinger equation for these $\ket{\Psi}$ and $H$ leads to a set of equations for $c_{e}(t)$ and $c_{g,\pmb{k}}(t)$:
\begin{equation}
    \dot{c}_{e}(t)=-i\sum_{\pmb{k}}g_{\pmb{k}}^{*}e^{i(\omega_{q}-\omega_{\pmb{k}})t}c_{g,\pmb{k}}(t)
\end{equation}
\begin{equation}
    \dot{c}_{g,\pmb{k}}(t)=-ig_{\pmb{k}}e^{-i(\omega_{q}-\omega_{\pmb{k}})t}c_{e}(t)
    \label{c_g}
\end{equation}
Reducing this set to only one equation for $c_{e}(t)$, one can write:
\begin{equation}
    \dot{c}_{e}(t)=-\sum_{\pmb{k}}|g_{\pmb{k}}|^{2}\int_{0}^{t}dt^{\prime}e^{i(\omega_{q}-\omega_{\pmb{k}})(t-t^{\prime})}c_{e}(t^{\prime})
    \label{c_e}
\end{equation}
Up to this moment the environmental properties have not played their role yet. The next step - changing the sum with the integral - has to take the 3D waveguide into account:
\begin{equation}
    \sum_{\pmb{k}}\rightarrow2\frac{L}{2\pi}\int_{0}^{\infty}dk.
    \label{transition}
\end{equation}
The factor 2 here comes from two possible directions of $\pmb{k}$, while there is only one possible polarisation direction, as we consider only decay into the fundamental mode of the waveguide. $L$ is a "quantization length" playing a similar role in the waveguide as the quantization volume in an open 3D space. With respect to \ref{transition}, it is important to notice that we understand $k=|\pmb{k}|$ as an effective wavenumber in the waveguide, i.e. as $\sqrt{{k}_{r}^{2}-{k}_{c}^{2}}$, where ${k}_{r}$ is the wavenumber in free space, and $k_{c}=\sqrt{(\frac{m\pi}{a})^2+(\frac{n\pi}{b})^2}$ is the cutoff wavenumber of the rectangular waveguide with cross section dimensions $a$ and $b$. As we are interested only in the first TE spatial mode ($m$=1 and $n$=0, providing $a>b$), in our case $k_{c}=\frac{\pi}{a}$.
\begin{figure}[ht]
\centering
\includegraphics[width=0.5\linewidth]{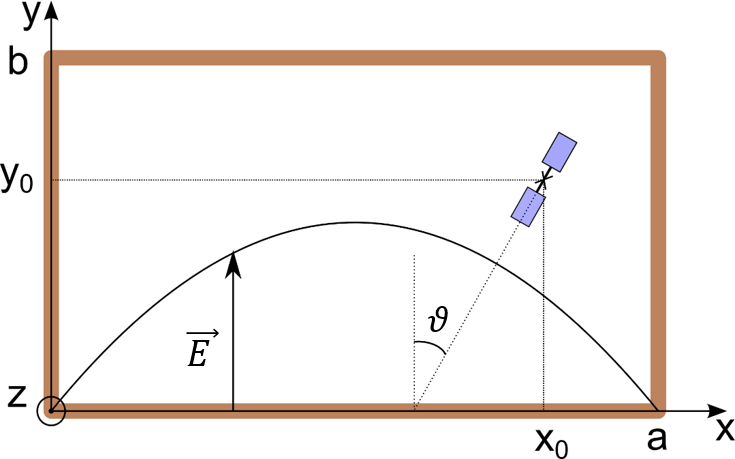}
\caption{Transmon qubit, considered as a point-like dipole, inside a rectangular waveguide with cross section dimensions $a$~x~$b$.}
\label{dip_in_wg}
\end{figure}
To move further we have to assume a specific model of interaction between the field and the emitter; for our experimental situation an electric dipole approximation is reasonable. Therefore we define atom-field coupling as $g_{\pmb{k}}=\frac{(\pmb{E_{\pmb{k}}}\pmb{d})}{\hbar}$ and use expression for an electric field amplitude created by a photon with frequency $\omega_{\pmb{k}}$ in a specific location $x_{0}$ inside the waveguide (see figure \ref{dip_in_wg})  $E_{\pmb{k}}=(\frac{\hbar\omega_{\pmb{k}}}{\epsilon V})^{\frac{1}{2}}\cdot sin(\frac{x_{0}\pi}{a})$, where $V=abL$ is the quantization volume. This brings us to:
\begin{equation}
    g_{\pmb{k}}=\sqrt{\frac{\omega_{\pmb{k}}}{\hbar\epsilon V}}|\pmb{d}_{eg}|cos(\vartheta)sin(\frac{x_{0}\pi}{a}).
    \label{g}
\end{equation}
with $\pmb{d}_{eg}$ being the transitional dipole moment of the atom and $\theta$ is the angle between $\pmb{E_{k}}$ and $\pmb{d}_{eg}$.

Using \ref{g} and \ref{transition} we rewrite \ref{c_e}:
\begin{equation}
    \dot{c}_{e}(t)=-2\frac{L}{2\pi}\int_{0}^{\infty}|\pmb{d}_{eg}|^2cos^{2}(\vartheta)sin^{2}(\frac{x_{0}\pi}{a})\frac{\omega_{\pmb{k}}}{\hbar\epsilon V}dk\cdot\int_{0}^{t}dt^{\prime}e^{i(\omega_{q}-\omega_{\pmb{k}})(t-t^{\prime})}c_{e}t^{\prime}.
    \label{c_e_2}
\end{equation}
 Using the dispersion relation in the waveguide $k=\sqrt{\epsilon\mu\omega^{2}_{\pmb{k}}-k^{2}_{c}}$ we switch to frequencies instead of wavenumbers:
\begin{equation}
    \dot{c}_{e}(t)=-\frac{\mu|\pmb{d}_{eg}|^{2}cos^{2}(\vartheta)sin^{2}(\frac{x_{0}\pi}{a})}{\pi\hbar ab}\cdot\int_{\omega_{c}}^{\infty}\frac{\omega^{2}_{\pmb{k}}d\omega_{\pmb{k}}}{\sqrt{\epsilon\mu\omega^{2}_{\pmb{k}}-k_{c}^{2}}}e^{i(\omega_{q}-\omega_{\pmb{k}})(t-t^{\prime})}
    \cdot\int_{0}^{t}c_{e}(t^{\prime})dt^{\prime},
\label{c_e_3}
\end{equation}
where $\omega_{c}=k_{c}/\sqrt{\epsilon\mu}$ is the cutoff frequency of the waveguide.

At this moment we make two other assumptions which are often referred to as "Wigner-Weisskopf approximation". Both assumptions are based on the idea that at any time during the decay only frequencies close to $\omega_{q}$ can be populated. With this in mind, we exchange in \ref{c_e_3} $\omega^{2}_{\pmb{k}}$ with $\omega^{2}_q$ and extend the lower integration limit to $-\infty$. This means that $\omega_{q}$ is sufficiently higher than $\omega_{c}$, so that in the frequency range we are interested in $\frac{\omega^{2}_{\pmb{k}}}{\sqrt{\epsilon\mu\omega^{2}_{\pmb{k}}-k_{c}^{2}}}$ can be assumed to be constant. In this case \ref{c_e_3} becomes:
\begin{equation}
    \dot{c}_{e}(t)=-\frac{\mu|\pmb{d}_{eg}|^{2}cos^{2}(\vartheta)sin^{2}(\frac{x_{0}\pi}{a})}{\pi\hbar ab}\frac{\omega^{2}_{q}}{\sqrt{\epsilon\mu\omega^{2}_{q}-k_{c}^{2}}}\cdot\int_{-\infty}^{\infty}d\omega_{\pmb{k}}e^{i(\omega_{q}-\omega_{\pmb{k}})(t-t^{\prime})}\int_{0}^{t}c_{e}(t^{\prime})dt^{\prime}.
\label{c_e_4}
\end{equation}
Using one of the Dirac delta function definitions $2\pi\delta(t^\prime-t)=\int_{-\infty}^{\infty}d\omega_{\pmb{k}}e^{i(\omega_{q}-\omega_{\pmb{k}})(t-t^{\prime})}$ and after integration over $t^\prime$ we finally arrive at the usual exponential decay:
\begin{equation}
    \dot{c}_{e}(t)=-\frac{\gamma}{2}c_{e}(t)
\label{c_e_5}
\end{equation}
\begin{equation}
    \gamma=\frac{2\mu|\pmb{d}_{eg}|^{2}cos^{2}(\vartheta)sin^{2}(\frac{x_{0}\pi}{a})}{\hbar ab}\frac{\omega^{2}_{q}}{\sqrt{\epsilon\mu\omega^{2}_{q}-k_{c}^{2}}}=\frac{2|\pmb{d}_{eg}|^{2}cos^{2}(\vartheta)sin^{2}(\frac{x_{0}\pi}{a})}{\hbar ab}\frac{\omega^{2}_{q}\cdot Z_{r}}{\sqrt{\omega^{2}_{q}-\omega_{c}^{2}}}.
\label{gamma}
\end{equation}
$Z_{r}=\sqrt{\frac{\mu}{\epsilon}}$ in the last expression is the impedance of the waveguide medium. Plugging the usual exponential solution of \ref{c_e_5} into \ref{c_g}, one can arrive at the Wigner-Weisskopf picture from the equations \ref{eqn:WW1}-\ref{eqn:WW2} from the main text. Therefore, providing all the aforementioned approximations are correct, one should expect the typical Wigner-Weisskopf picture of exponential decay to be accurate in the 3D waveguide as well, apart of expression for $\gamma$. Note, that for the atomic frequencies $\omega_{q}$ close to the cutoff $\omega_{c}$ the assumptions of the Wigner-Weisskopf approximation are not correct and the decay is not exponential. For the experiment, described in the main text we are detecting frequency components of the decay within a 60~MHz frequency window around the qubit frequency 7.3~GHz. Within this window the factor $\frac{\omega^{2}_{\pmb{k}}}{\sqrt{\epsilon\mu\omega^{2}_{\pmb{k}}-k_{c}^{2}}}$ changes around 1.7\% which is small enough not to affect the decay.
\section{Methods}
\label{app_B}
In the experiment we use a standard 12cm long WR90 copper rectangular waveguide with cross section dimensions 10.16\,mm and 22.86\,mm and cutoff frequency 6.557\,GHz. On both ends it is connected to 50$\Omega$ coaxial cables with cable-to-waveguide adapters. Superconducting coils placed on the external surface of the waveguide were used to tune the qubits frequencies. The DC lines of the coils were heavily filtered to avoid additional flux noise at the qubit locations, including home-made steel powder filters \cite{lukashenko_improved_2008} installed on the top of the shielding can. The waveguide was thermally anchored to the baseplate of a dilution refrigerator and operated at $\sim$20\,mK. Qubits were inserted into the waveguide through specially designed slits, and mechanically fixed with copper clamps, allowing for appropriate thermalization. Appropriate RF and DC shielding of the sample was provided by a double layer cryoperm shield, and a niobium shield.

For qubit fabrication we used a standard bridge-free Al/AlO$_{x}$/Al technique, described in details elsewhere \cite{lecocq_junction_2011}. In short, electron beam lithography was used to pattern junctions and antennas on a 2-inch 330\,$\mu$m thick commercial sapphire substrate. An electron-gun evaporator was used to deposit 2 layers of Al on the substrate with a thickness of 20 and 30\,nm respectively. Josephson junctions were formed between the 2 layers by 2.5 minutes oxidation under 15 mbar pressure, resulting in a critical current density $\sim$50\,A/cm$^2$. To achieve appropriate balance between flux tunability and susceptibility to the flux noise, both qubits had nonidentical junctions. The sapphire wafer was subsequently diced with a diamond saw, to 2.5\,mm-wide and 2.5\,cm-long chips.
\section{Experimental setup}
\label{app_C}
\begin{figure}[ht]
\centering
\includegraphics[width=0.5\linewidth]{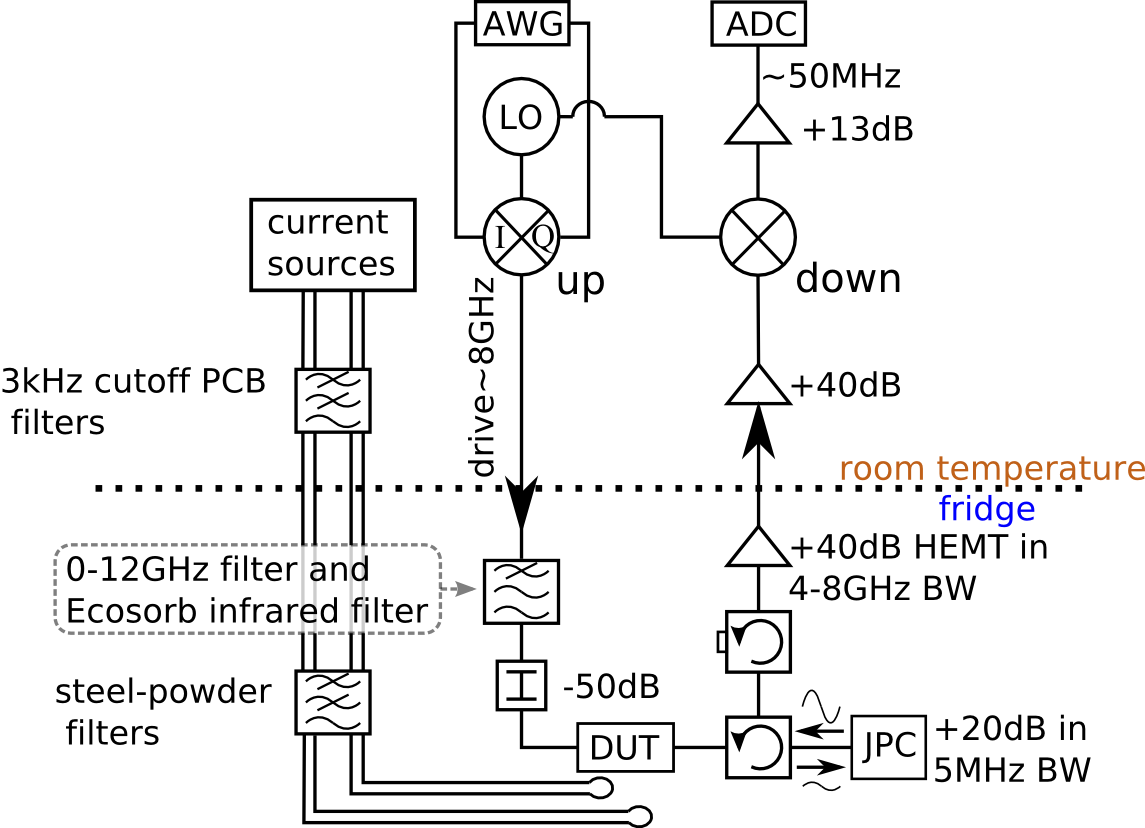}
\caption{Experimental setup.}
\label{exp_setup}
\end{figure}

The experimental setup is shown in figure \ref{exp_setup}. We are generating our pulses at low ($\sim$200\,MHz) frequency from 2 independent channels of an Arbitrary Waveform Generator and up-convert them to the qubit frequency with an IQ-mixer. The pulses are attenuated by 50\,dB  before entering the waveguide installed at the baseplate of the fridge. After interaction with the qubits, the signal is first amplified by 20\,dB within a narrow frequency band ($\sim$5\,MHz) at the base plate of the refrigerator using a JPC and consequently at the 4K stage by a High Electron Mobility Transistor amplifier (+40\,dB) and at room temperature by another low-noise amplifier (also +40\,dB). The input line of the fridge was equipped with a home-made Eccosorb infrared filter as well as a 12\,GHz cutoff low-pass filter (both are not shown). We use a conventional mixer for down converting the signal to 50\,MHz and an analog-to-digital converter for digitizing and averaging the signal amplitude. The effective noise temperature without and with JPC was estimated to be about 10.5\,K and 80\,mK respectively.

Superconducting coils, used to tune the qubit frequencies, were powered from current sources through a set of low-pass filters: 1.9\,MHz cutoff, homemade 500\,kHz and 3\,kHz cutoff resistive Pi-filters and steel powder filters with a $\sim$30\,kHz cutoff. The filters were installed inside an aluminum housing on the base plate of the refrigerator outside the shielding-can of the sample.
\section{Theoretical model: "Infinitely long" pulse}
\label{app_D}
The next two appendices describe the model, based on input-output approach, used to fit the experimental data from the main text. This subsection is devoted to the situation in the main text referred to as "infinitely long" pulse, where everything can be derived analytically. In the next subsection we describe the model we solve numerically for arbitrary length pulses.

Let us consider a driven 2-level system in the framework of optical Bloch equations and input-output theory:
\begin{equation}
\begin{split}
    &\frac{\mathrm{d}}{\mathrm{dt}} \left< \sigma^z(t) \right> =- 2\mathrm{i}\sqrt{\frac{\gamma\beta}{2}}\Bigl[ \alpha e^{-\mathrm{i}\omega t}\left<\sigma^+(t)\right>- \alpha^{*} e^{\mathrm{i}\omega t}\left<\sigma^-(t)\right>\Bigr]- \gamma \left( \left< \sigma^z(t) \right> + 1\right)\\
    &\frac{\mathrm{d}}{\mathrm{dt}} \left< \sigma^-(t) \right> = -\left( \mathrm{i}\omega_{q} + \frac{\gamma}{2}\right)\left<\sigma^-(t)\right> + \mathrm{i}\alpha e^{-\mathrm{i}\omega t}\sqrt{\frac{\gamma\beta}{2}} \left< \sigma^z(t) \right>\\
    &\left<a_{out}\right> = \left<a_{in}\right> - \mathrm{i}\sqrt{\frac{\gamma\beta}{2}} \left< \sigma^-(t) \right>,
\end{split}
\end{equation}
where $\omega_{q}$ is the qubit transition frequency, $\gamma$ is the qubit linewidth, related to both, spontaneous emission into the waveguide and internal losses with $\gamma=\gamma_{waveguide}+\gamma_{internal}$), $\alpha$ is the drive amplitude and $\beta=\gamma_{waveguide}/(\gamma_{waveguide}+\gamma_{internal})$ is a parameter characterising the portion of the qubit energy emitted into the waveguide.

Switching into the rotating (with the drive) frame and changing notation for $\left< \sigma^z(t)\right>$:
\begin{equation}
    \left< \sigma^-(t) \right> = e^{-\mathrm{i}\omega t}s\textnormal{, }\left< \sigma^+(t) \right> = e^{\mathrm{i}\omega t}s^{*}\textnormal{, }\left< \sigma^z(t) \right> = 2\rho - 1,
\end{equation}
where $s$ represents the coherence and $\rho$ is the excited population of the qubit. Both are changing slowly in comparison with $\omega$. This yields:
\begin{equation}
    \begin{split}
        &\frac{\mathrm{d}}{\mathrm{dt}} \rho =- \mathrm{i}\sqrt{\frac{\gamma\beta}{2}}\left[ \alpha s^{*} - \alpha^{*}s \right] - \gamma \rho\\
        &\frac{\mathrm{d}}{\mathrm{dt}} s = -\left[ \mathrm{i}\left( \omega_{q} -\omega\right) + \frac{\gamma}{2}\right]s + \mathrm{i}\alpha \sqrt{\frac{\gamma\beta}{2}} \left(2\rho-1\right)
    \end{split}
\end{equation}

which for steady state reduces to:
\begin{equation}
    s_{0} = \frac{\mathrm{i} \alpha \sqrt{\frac{\gamma \beta}{2}}}{\mathrm{i}\left( \omega_{q} -\omega\right)+ \frac{\gamma}{2}}\left( 2\rho_{0} - 1\right),
\end{equation}
and:
\begin{equation}
    \rho_{0}= -\mathrm{i}\sqrt{\frac{\beta}{2\gamma}} \left( \alpha s_{0}^{*} - \alpha^{*} s_{0} \right)
\end{equation}
Choosing the initial phase so $\alpha$ is real, we finally obtain:
\begin{equation}
\begin{split}
    & s_{0} = -\frac{\alpha \sqrt{\gamma\beta/2}\left( \omega_{q} - \omega +\mathrm{i}\gamma/2\right)}{\left( \omega_{q} - \omega \right)^2 + \gamma^2/4 +\alpha^2\beta \gamma }\\
    &\rho_{0} = \frac{\alpha^2\beta \gamma/2}{\left( \omega_{q} - \omega \right)^2 + \gamma^2/4 +\alpha^2\beta \gamma }
\end{split}
\end{equation}
We are going to use the obtained steady state solution as an initial condition to fit the qubit decay without drive. The drive is abruptly switched off at a given time $t=0$. The system then evolves according to:
\begin{equation}
\begin{split}
    &\frac{\mathrm{d}}{\mathrm{dt}} \left< \sigma^z(t) \right> = - \gamma \left( \left< \sigma^z(t) \right> + 1\right)\\
    &\frac{\mathrm{d}}{\mathrm{dt}} \left< \sigma^-(t) \right> = -\left( \mathrm{i}\omega_{q} + \frac{\gamma}{2}\right)\left<\sigma^-(t)\right>\\
    &\left<a_{out}\right> = - \mathrm{i}\sqrt{\frac{\gamma\beta}{2}} \left< \sigma^-(t) \right>,
\end{split}
\end{equation}
This leads to the usual exponential decay with initial condition set by $s_{0}$ and $\rho_{0}$:
\begin{equation}
\begin{split}
    &\left< \sigma^z(t) \right> = 2e^{-\gamma t}\rho_{0} - 1\textnormal{, }\left< \sigma^-(t) \right> = e^{-\mathrm{i}\omega_{q} t - \gamma t /2}s_{0}\\
    &\left< a_{out}(t) \right> = -\sqrt{\frac{\gamma\beta}{2}} e^{-\mathrm{i}\omega_{q} t - \gamma t /2}s_{0}
\end{split}
\end{equation}
Now we have to add the amplifier to the picture:

\begin{equation}
    a_{amp}(t) = \sqrt{G} \int_{-\infty}^{t} \left<a_{out}(\tau)\right> e^{-\left[\mathrm{i}\omega_{amp}+\kappa\right]\left(t-\tau\right)}\mathrm{d}\tau +h,
    \label{filter_function}
\end{equation}
where $\omega_{amp}$ is the center frequency of the amplifier band $\kappa$, $G$ is the gain, while $h$ is a noise operator.

\begin{figure*}
\raggedright
\textbf{a}\hspace{3.3cm}\textbf{1 qubit}\hspace{4.4cm}\textbf{b}\hspace{3.3cm}\textbf{2 qubits}
\includegraphics[width=0.45\linewidth]{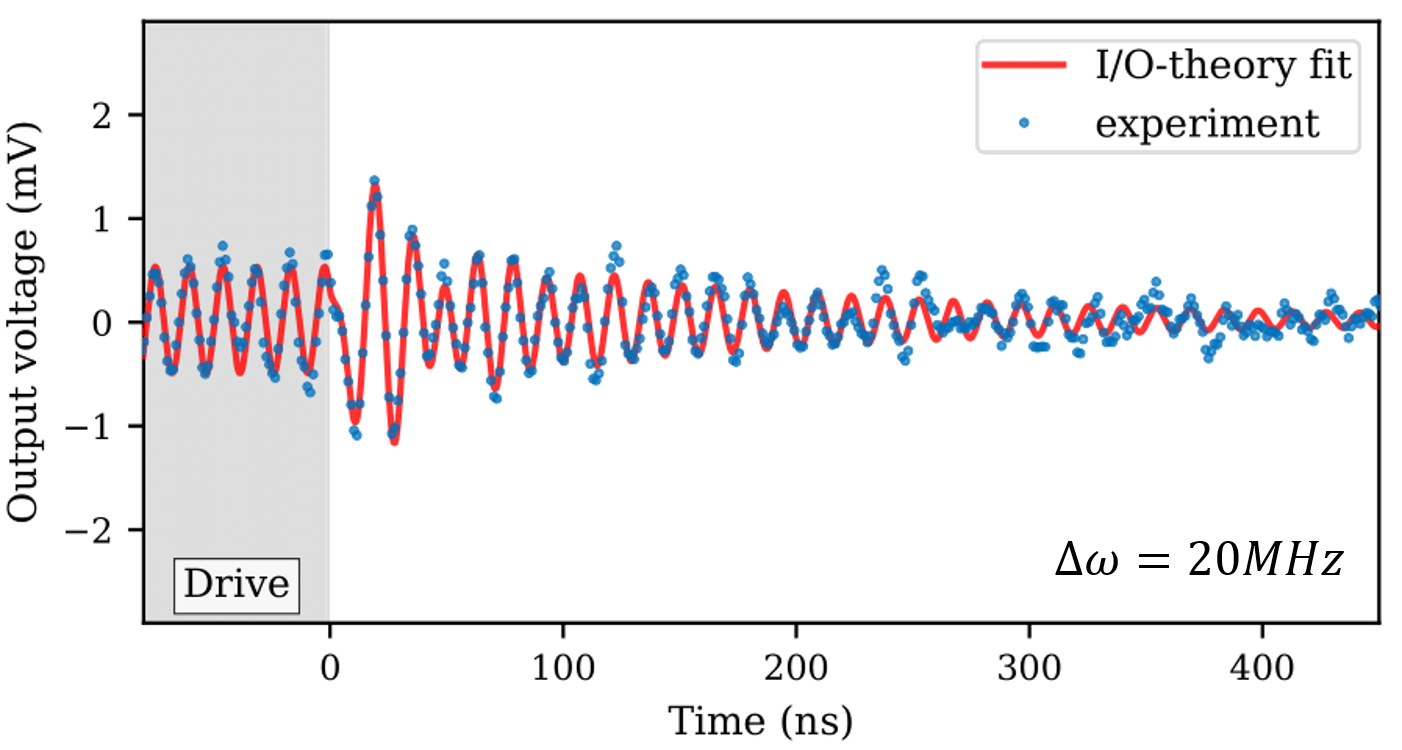}\hspace{1.5cm}\includegraphics[width=0.45\linewidth]{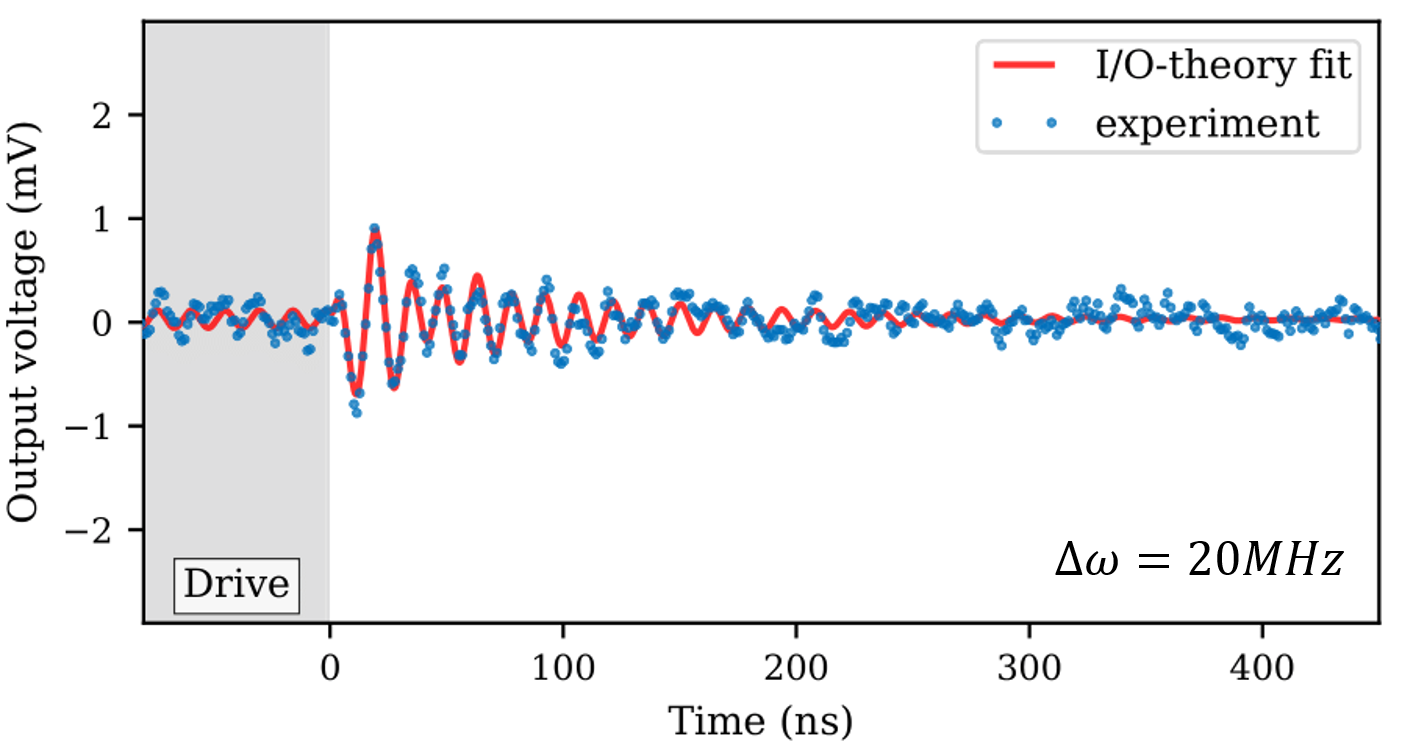}\par
\textbf{c}\hspace{8.9cm}\textbf{d}\par
\includegraphics[width=0.45\linewidth]{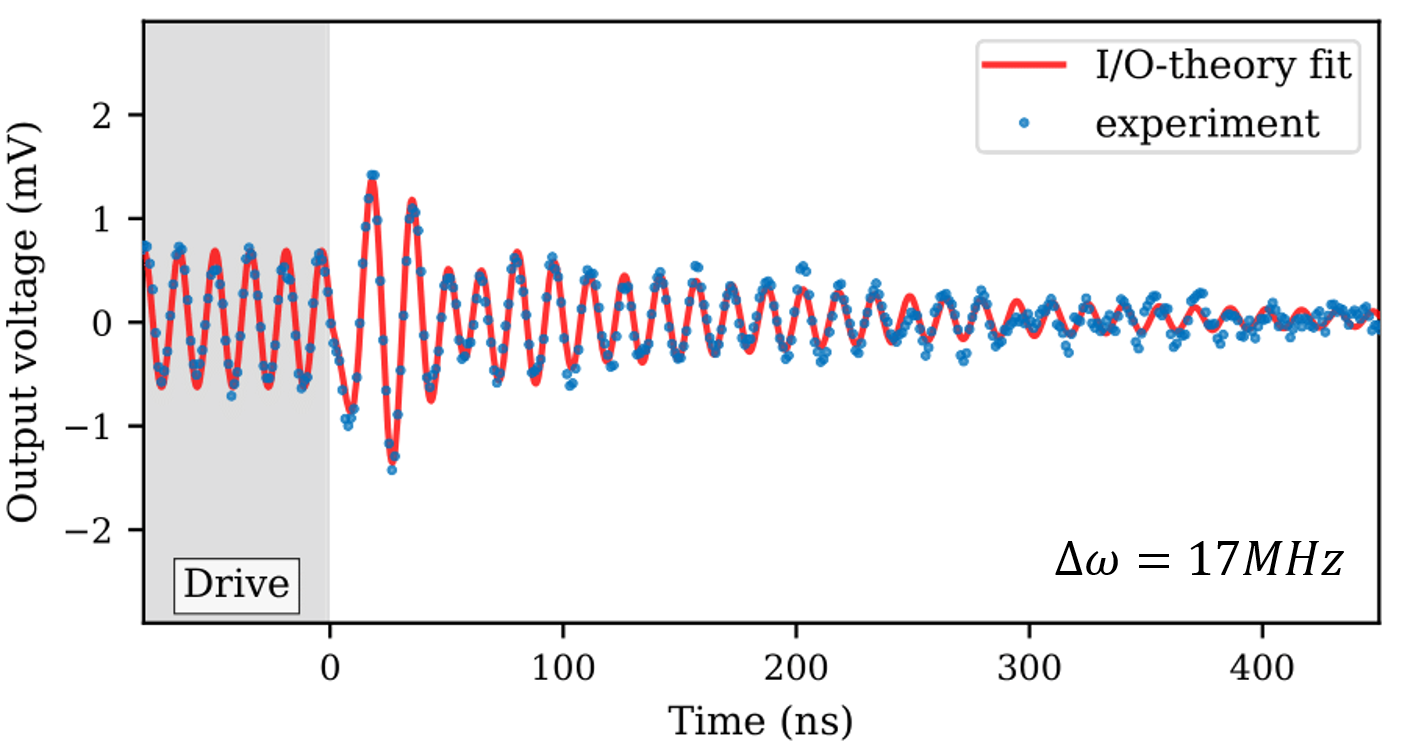}\hspace{1.5cm}\includegraphics[width=0.45\linewidth]{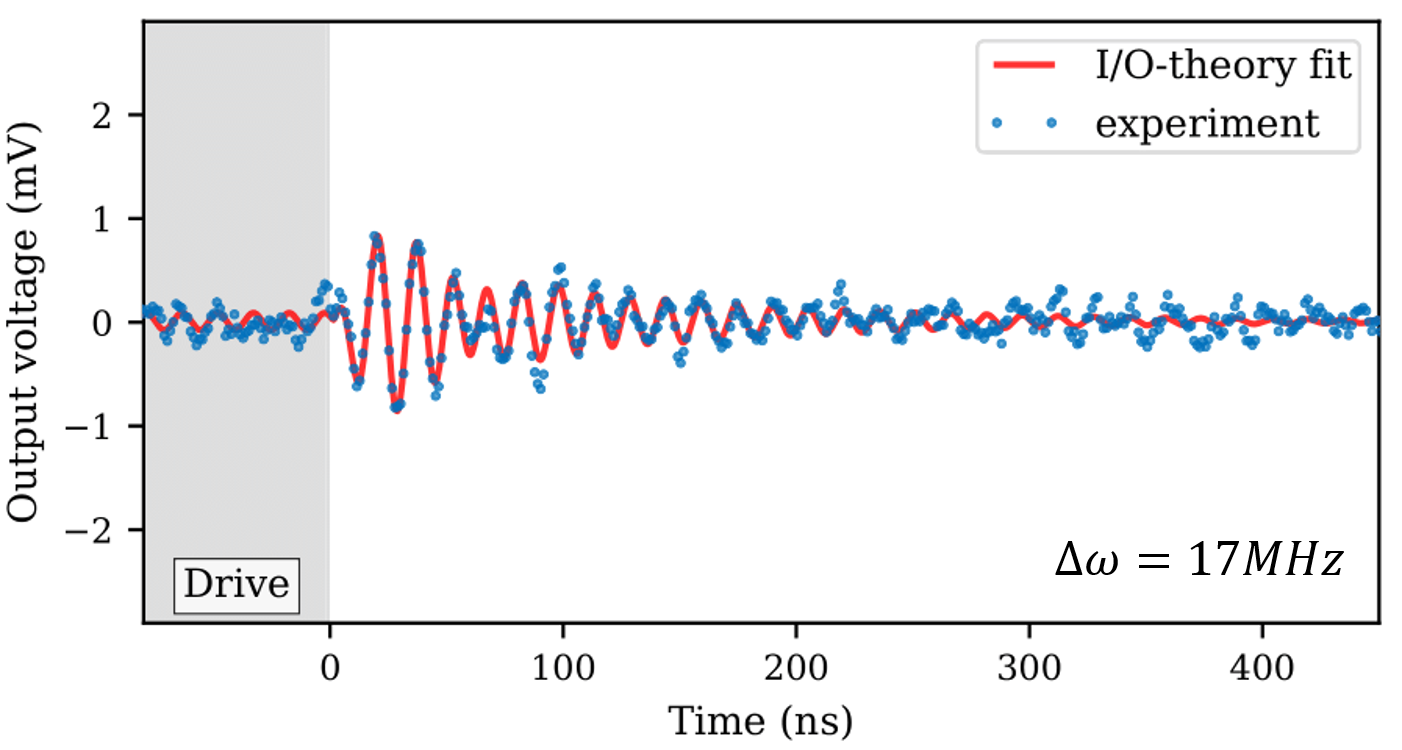}\par
\textbf{e}\hspace{8.9cm}\textbf{f}\par
\includegraphics[width=0.45\linewidth]{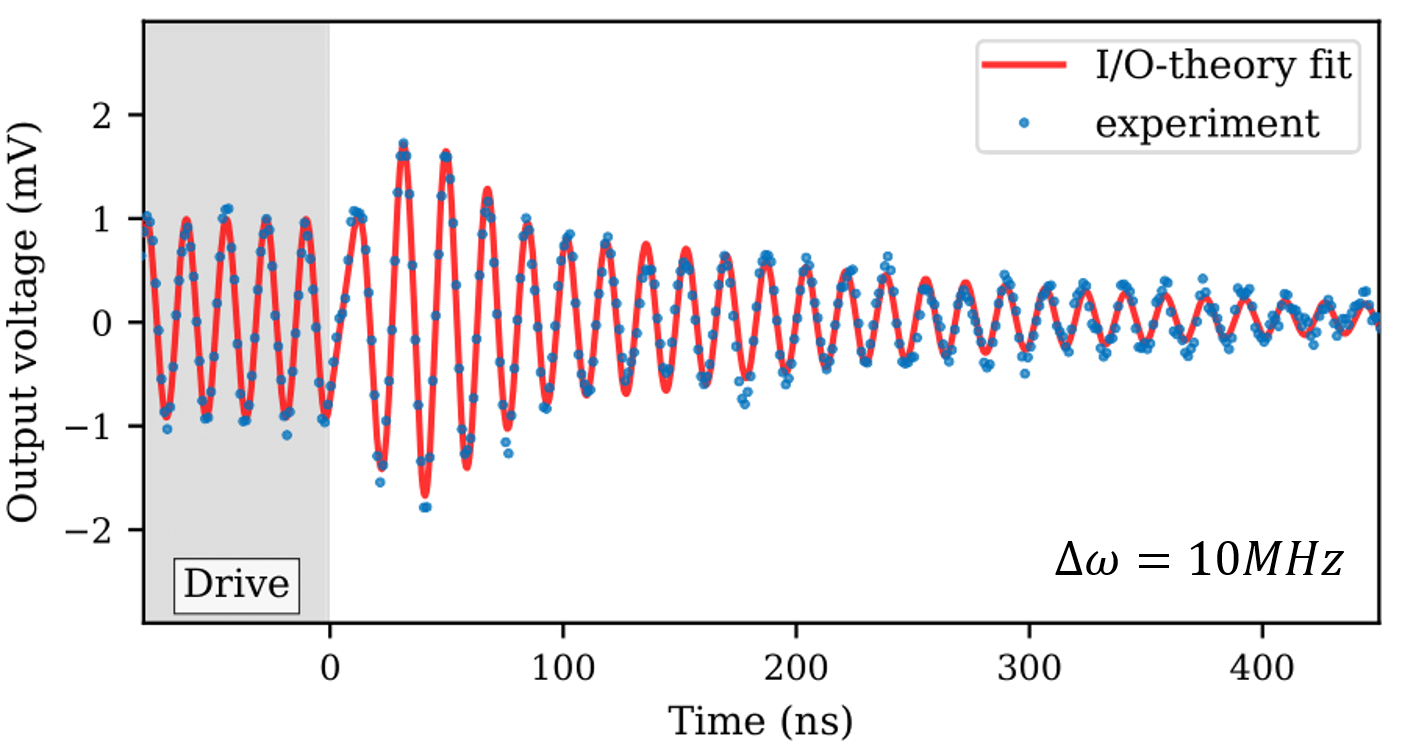}\hspace{1.5cm}\includegraphics[width=0.45\linewidth]{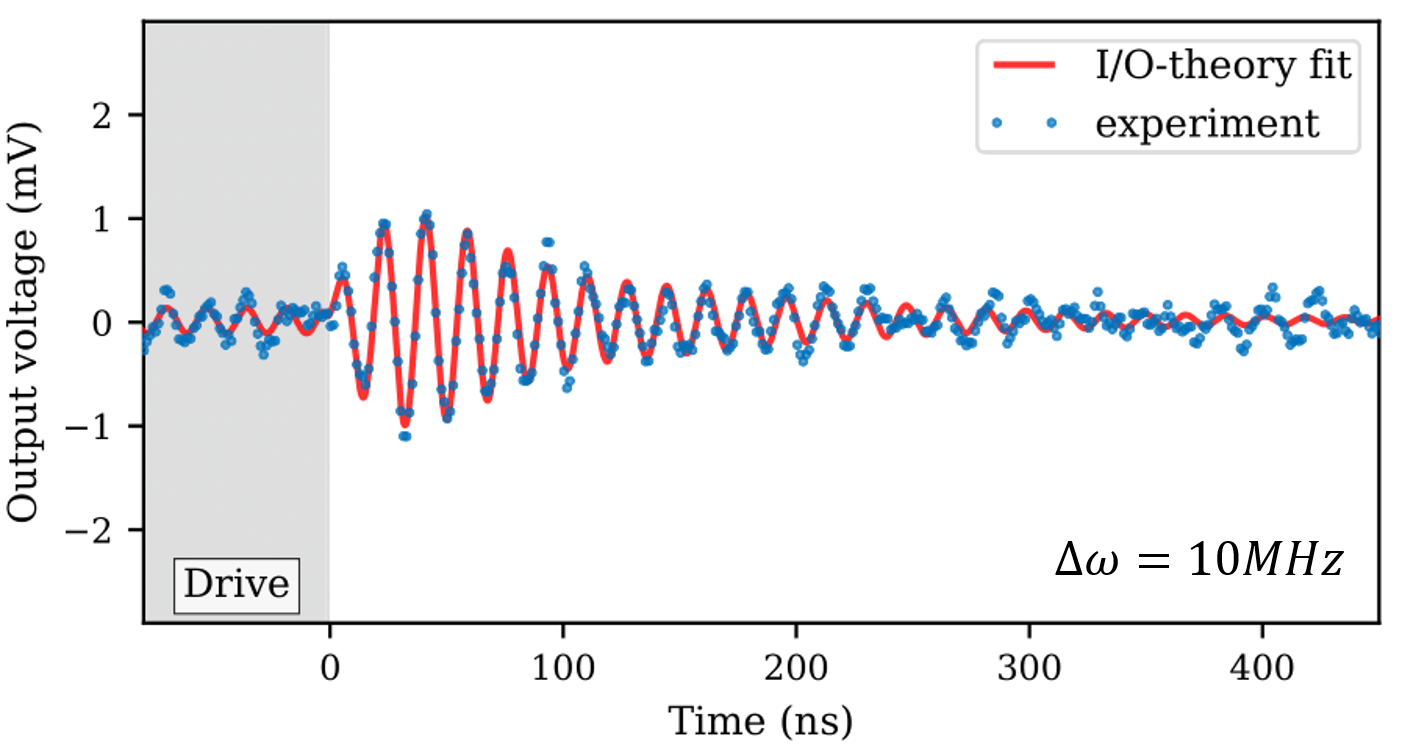}\par
\textbf{g}\hspace{8.9cm}\textbf{h}\par
\includegraphics[width=0.45\linewidth]{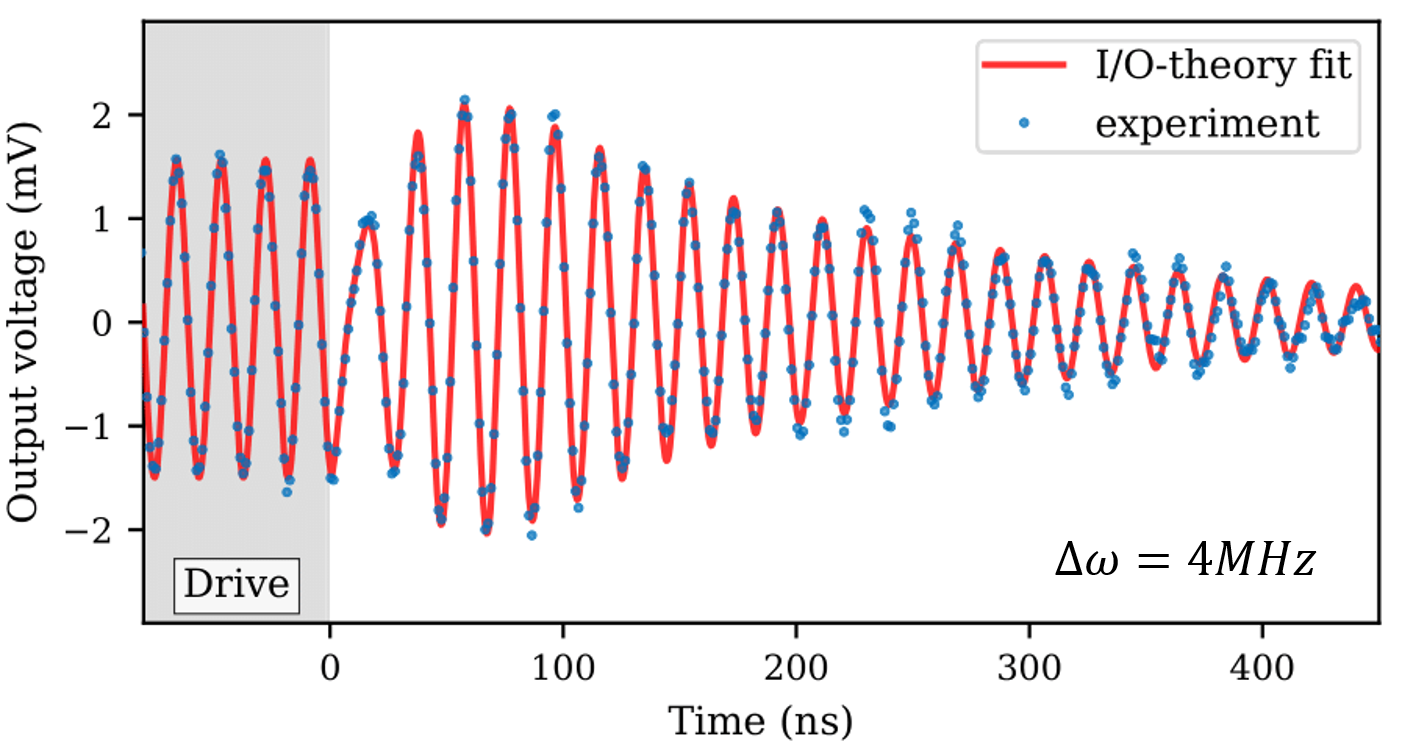}\hspace{1.5cm}\includegraphics[width=0.45\linewidth]{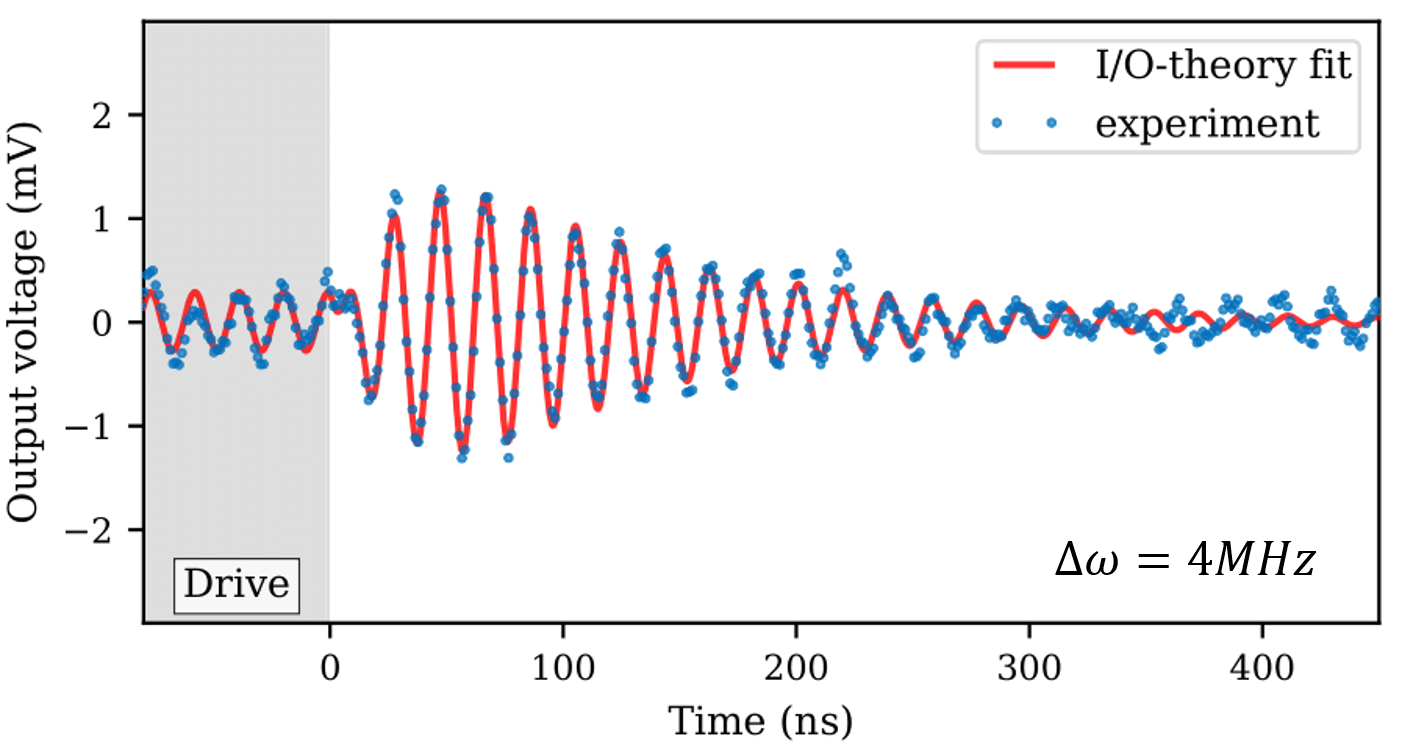}\par
\textbf{i}\hspace{8.9cm}\textbf{j}\par
\includegraphics[width=0.45\linewidth]{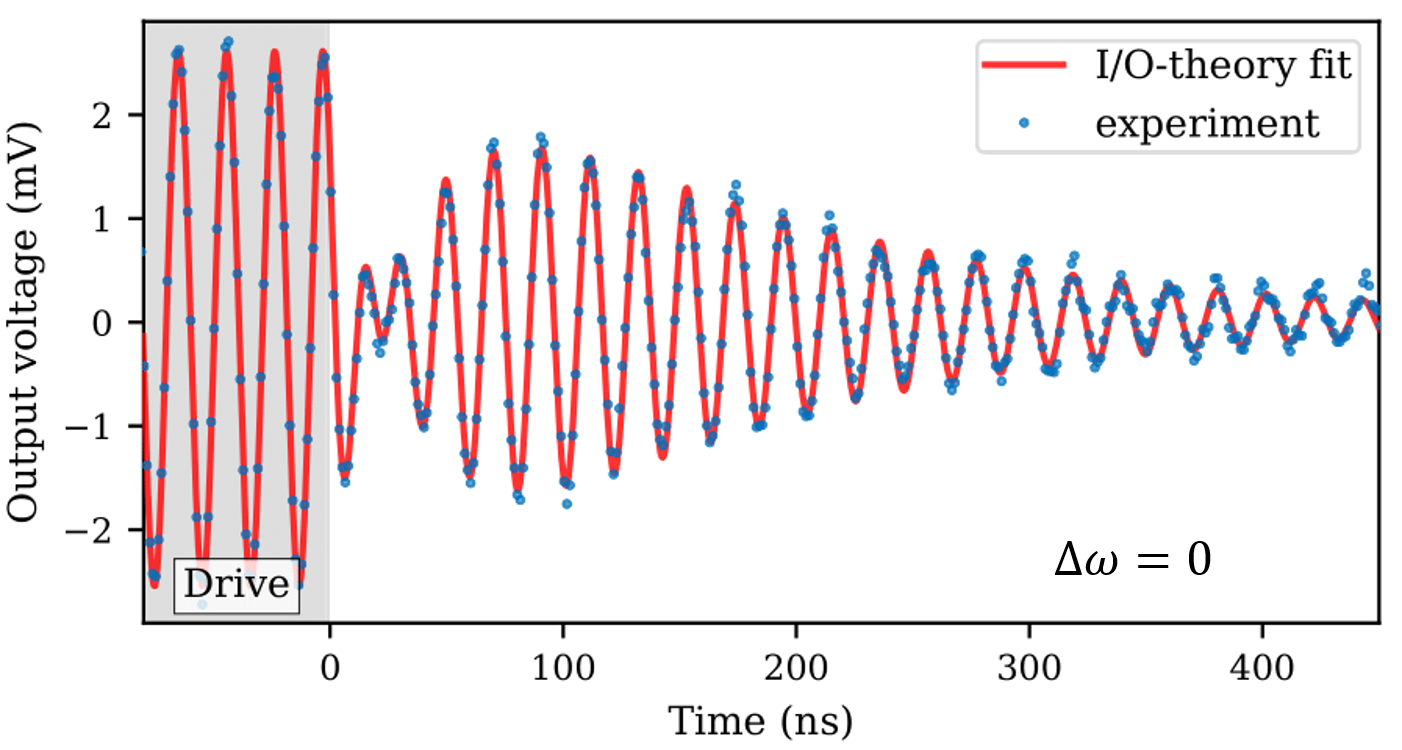}\hspace{1.5cm}\includegraphics[width=0.45\linewidth]{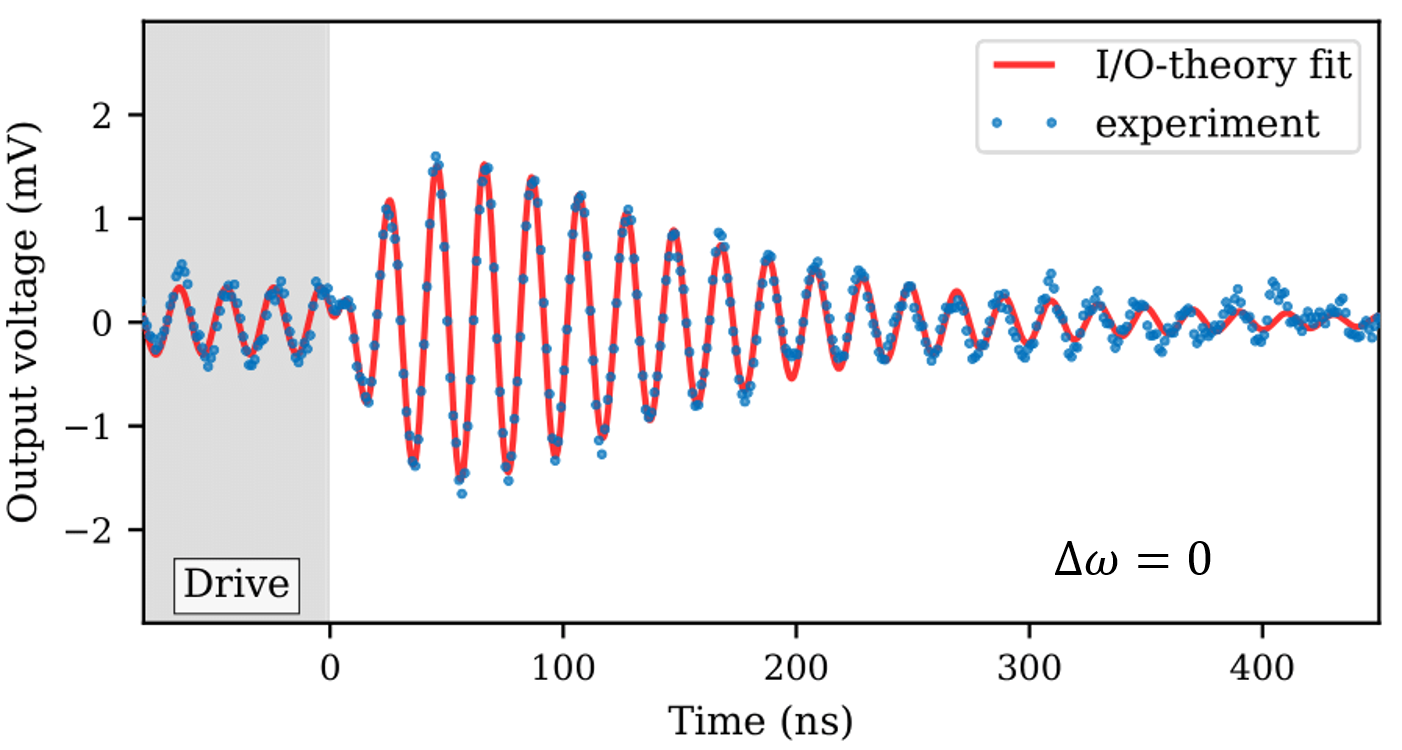}
\caption{Fitting experimental data for a single qubit (left column) and the bright state of 2 qubits (right column) for different detunings between drive and JPC frequencies $\Delta\omega=\omega-\omega_{amp}$. The data is taken from figures 4(a) and (c) from the main text.}
\label{fitting}
\end{figure*}

\begin{figure*}[!h]
\raggedright
\textbf{a}\hspace{8.9cm}\textbf{b}\par
\includegraphics[width=0.45\linewidth]{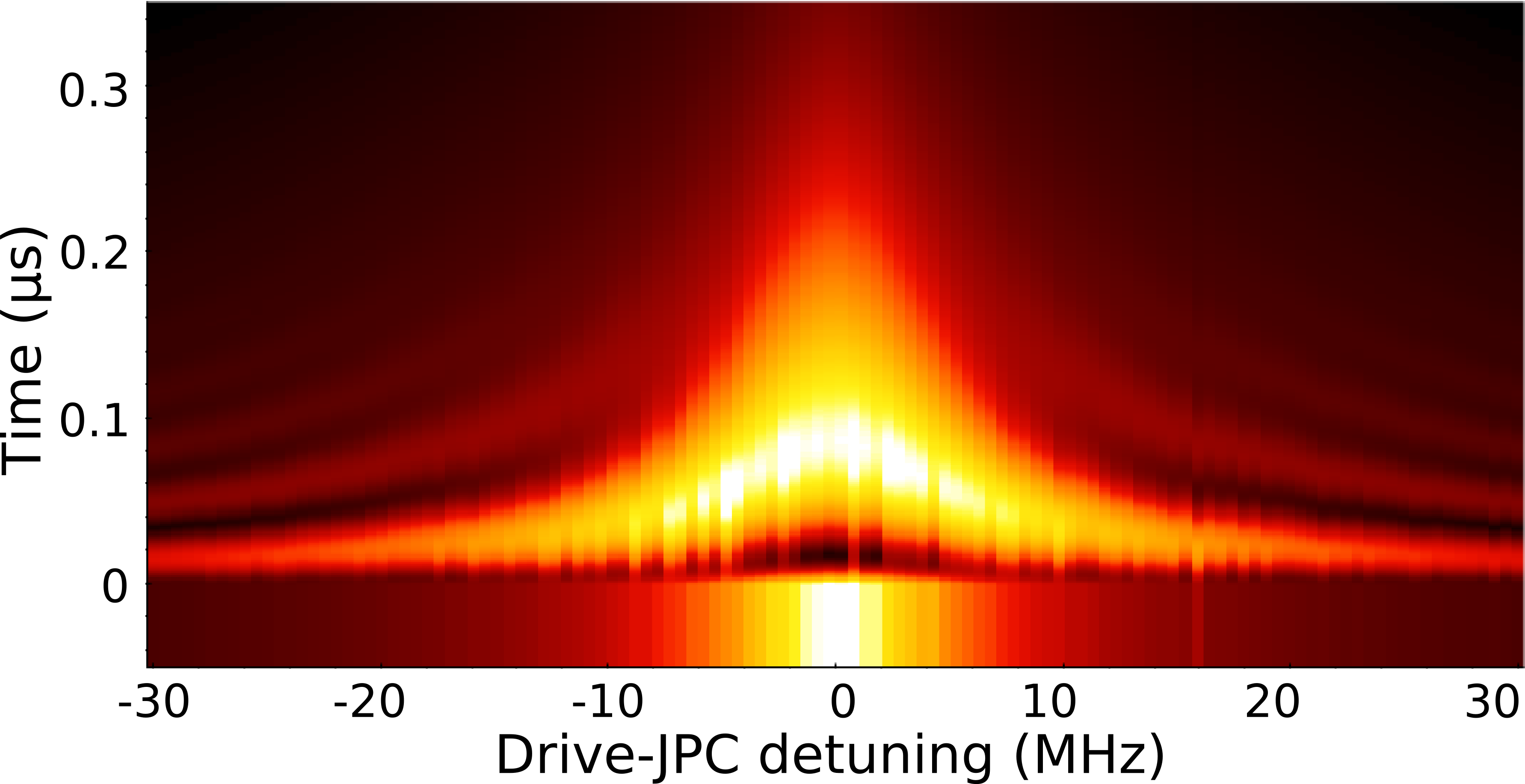}\hspace{1cm}
\includegraphics[width=0.45\linewidth]{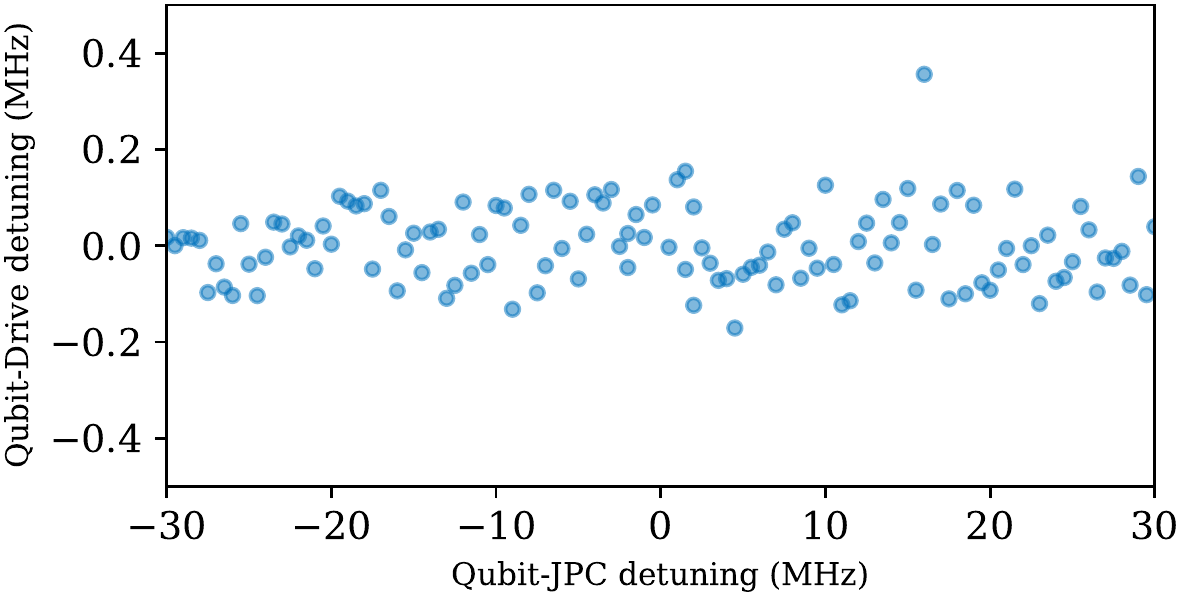}\par
\textbf{c}\hspace{8.9cm}\textbf{d}\par
\includegraphics[width=0.45\linewidth]{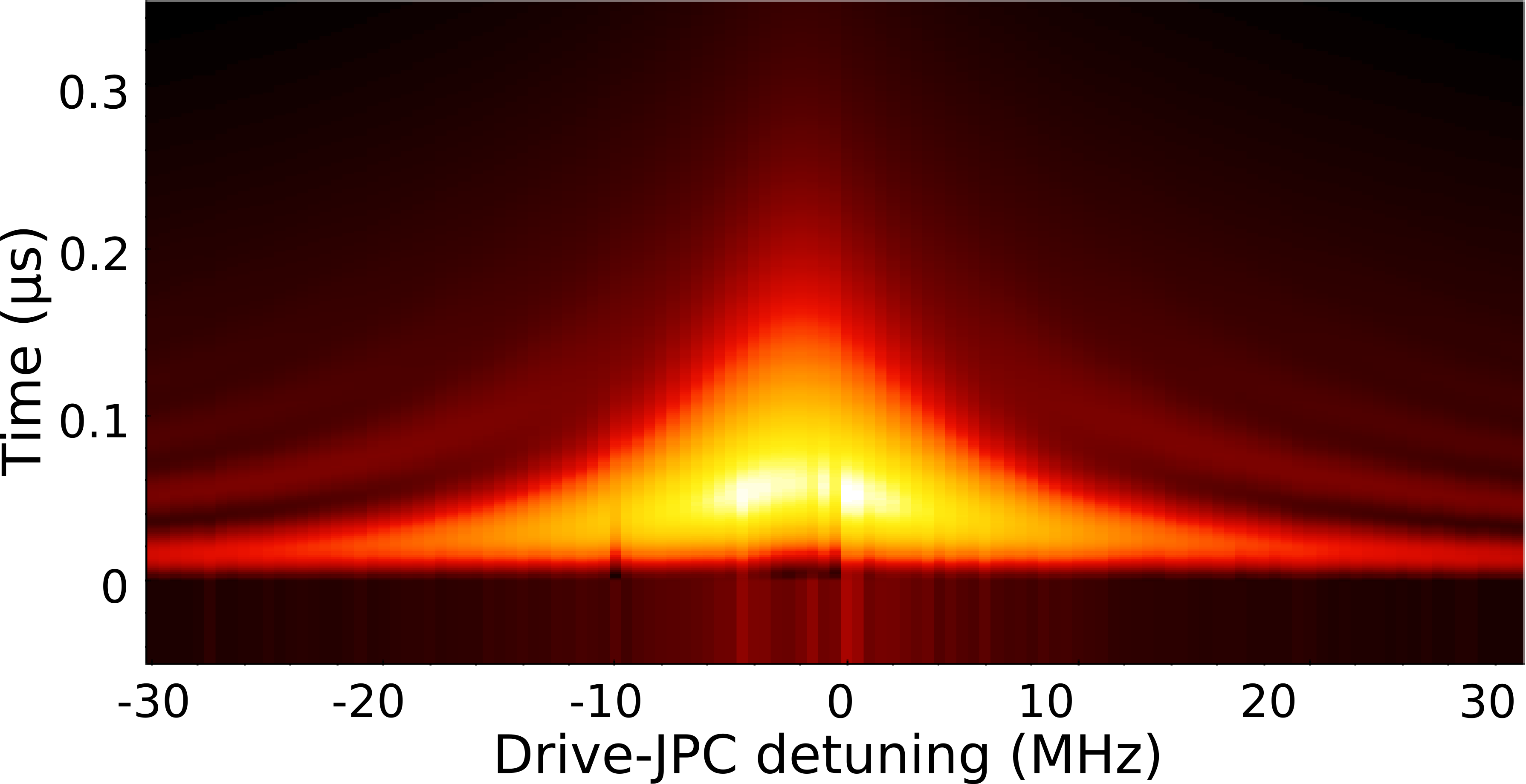}\hspace{1cm}
\includegraphics[width=0.45\linewidth]{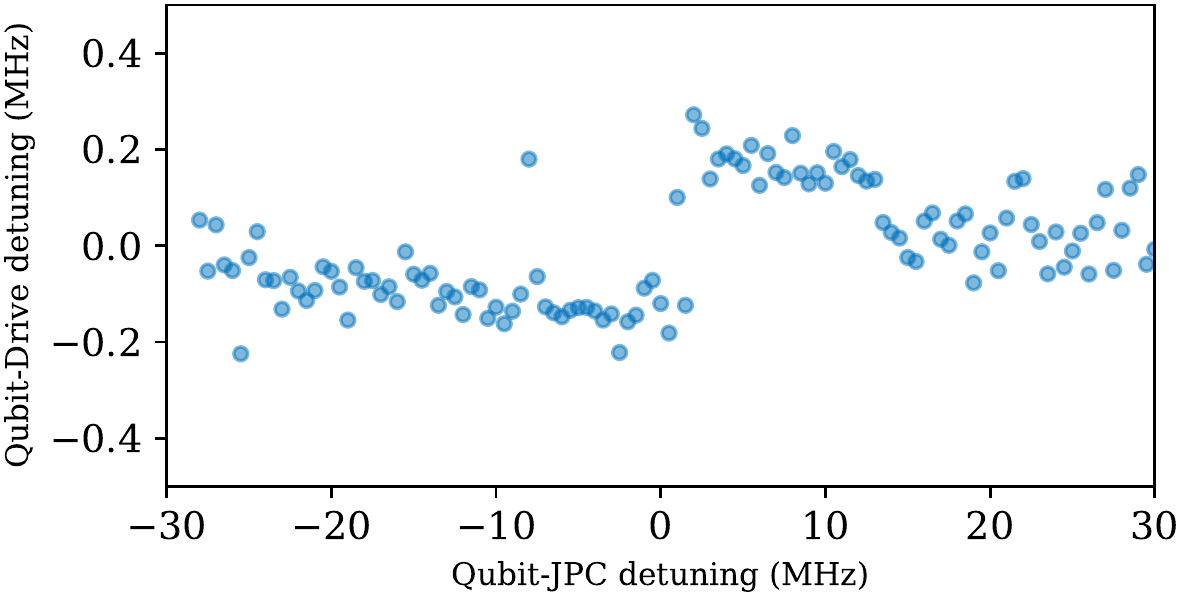}
\caption{\textbf{a} and \textbf{c}, Theoretical fits of the experimental data from figure \ref{fig:wf}(a) and (c) in the main text with the formula \ref{eqn:aamp}. Qubit-drive detuning is an independent fitting parameter for each drive-JPC detuning (unlike figures \ref{fig:wf}(b) and (d) where the qubit-drive detuning is always assumed to be zero); \textbf{b} and \textbf{d},  Qubit-drive detunings extracted from the fits for 1 and 2 qubits.}
\label{fitting-2}
\end{figure*}

For the transmitted field after the amplifier, we then obtain:
\begin{equation}
    \left<a_{amp}(t)\right>=\sqrt{G} \int_{-\infty}^t  e^{-\left[\mathrm{i}\omega_{amp}+\kappa\right]\left(t-\tau\right)} \Bigl[ e^{-\mathrm{i}\omega t}\left( \alpha-\mathrm{i} s_{0} \sqrt{\frac{\gamma\beta}{2}}\right)\Theta(-\tau)-s_{0}\sqrt{\frac{\gamma\beta}{2}} e^{-\mathrm{i}\omega_{q} t - \gamma t /2} \Theta(+\tau) \Bigr] \mathrm{d}\tau
\end{equation}
where $\Theta(t)$ is the Heaviside step function. This can be evaluated analytically and after some algebra one arrives at the final result:

\begin{equation}
    \begin{split}
        \left< a_{amp}(t)\right> = & \frac{e^{-t (\kappa + \mathrm{i} \omega_{amp}) + (\kappa - i (\omega - \omega_{amp})) \mathrm{Min}[0,t]} \left( \alpha - \frac{\alpha \beta \gamma/2 (\gamma/2 - \mathrm{i} (\omega_{q} - \omega)) }{ \alpha^2 \beta \gamma + \gamma^2/4 +(\omega_{q} - \omega)^2}\right)}{\kappa -\mathrm{i} (\omega - \omega_{amp})}+\\
        & + \frac{e^{-t (\kappa + \mathrm{i} \omega_{amp})} \left( -1 +e^{(-(\gamma/2) - \mathrm{i} \omega_{q} + \kappa +i \omega_{amp}) \mathrm{Max}[0,t]} \right) \alpha \beta \gamma/2 (\gamma/2 - \mathrm{i} (\omega_{q} - \omega))}{(\alpha^2 \beta \gamma + \gamma^2/4 + (\omega_{q} - \omega)^2) (\gamma/2 + \mathrm{i} (\omega_{q} + \mathrm{i} \kappa - \omega_{amp}))}.\label{eqn:aamp}
    \end{split}
\end{equation}
In this expression the first term describes the system behaviour at the times $t<0$ when the drive is on and the second term is 0. After the drive stops (at $t=0$), the second term describes the qubit decay (with the rate $\gamma/2$ instead of $\gamma$ as it corresponds to the amplitude reduction rate), while the first one describes the amplifier decay (with the rate $\kappa$). Formula \ref{eqn:aamp} was used to fit our experimental data for the from the figures 4(a) and (c) in the main text. Figure \ref{fitting} shows how well the fit works for the state of one qubit and the collective state of two qubits for different values of the drive-JPC detuning. In the experiment we intend to keep the drive frequency in resonance with the qubit. However, qubit frequency stability was noticeably affected by magnetic field noise in the cryostat. Since the theory does not require us to fix $\omega=\omega_{q}$ the qubit-drive detuning can be extracted as a result of the fit. Fitted experimental data for different detunings $\delta \omega$ for the one- and two-qubits cases, similar to figures 4(a) and (c) from the main text, are shown in figures \ref{fitting-2}(a) and \ref{fitting-2}(c). The qubit drive detuning extracted from the fit for every frequency is shown in figure \ref{fitting-2}(b), and mostly does not exceed 200\,kHz. Comparison of the figure \ref{fitting-2}(a) and 4(a) from the main text leads us to the conclusion that this detuning is the main source of noise for our experimental data.

The theory only describes the system dynamics in the case of the "infinitely long" pulse, after the dynamical equilibrium between the drive and qubit decay has been established. The case of short ($\pi/2$, $\pi$, $3\pi/2$...) pulses has been treated numerically and is described in the next subsection.
\section{Theoretical model: Arbitrary length pulses}
\label{app_E}
\begin{figure*}[!h]
\raggedright
\textbf{a}\hspace{8.9cm}\textbf{b}\par
\includegraphics[width=0.45\linewidth]{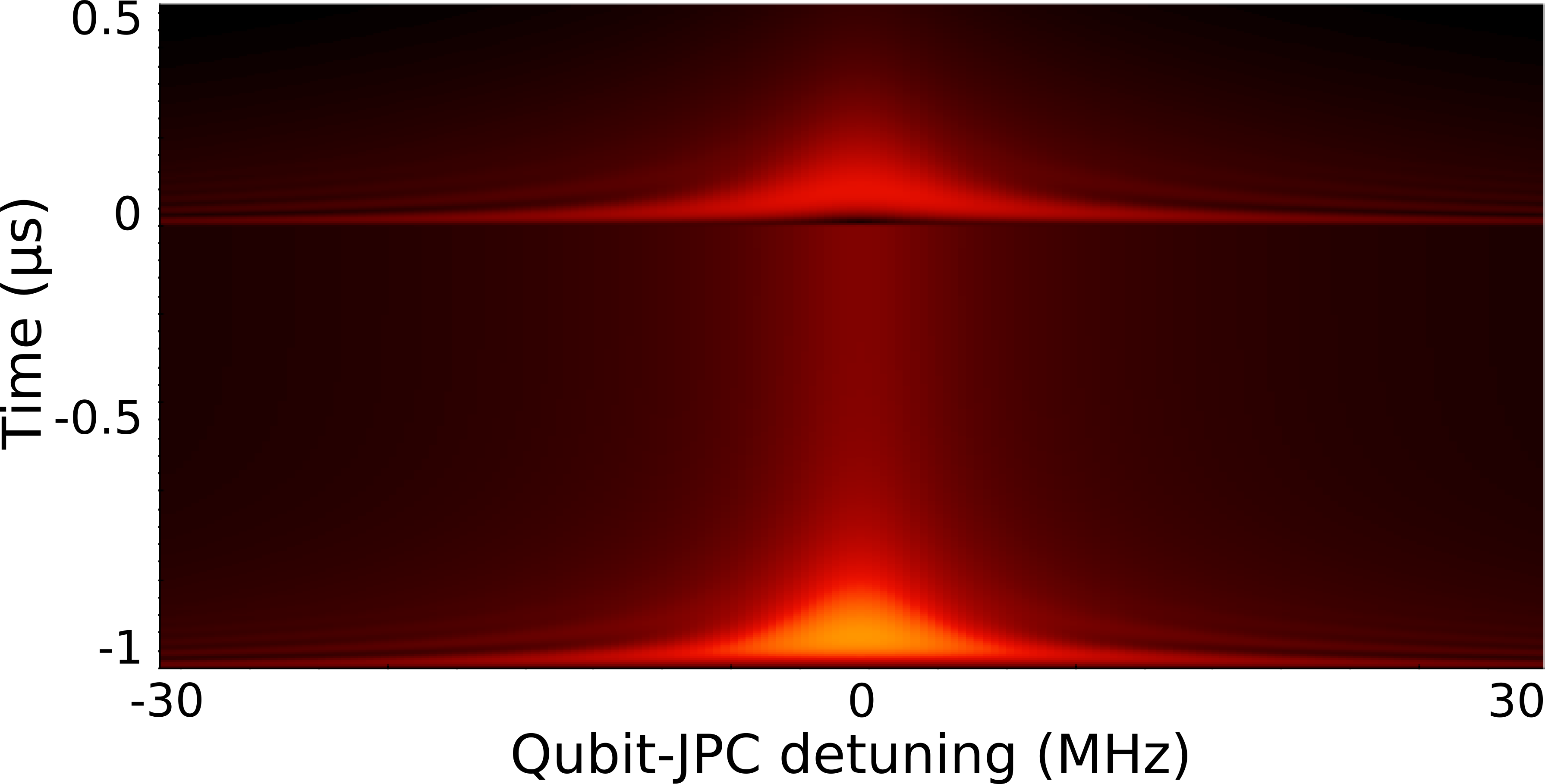}\hspace{1cm}
\includegraphics[width=0.45\linewidth]{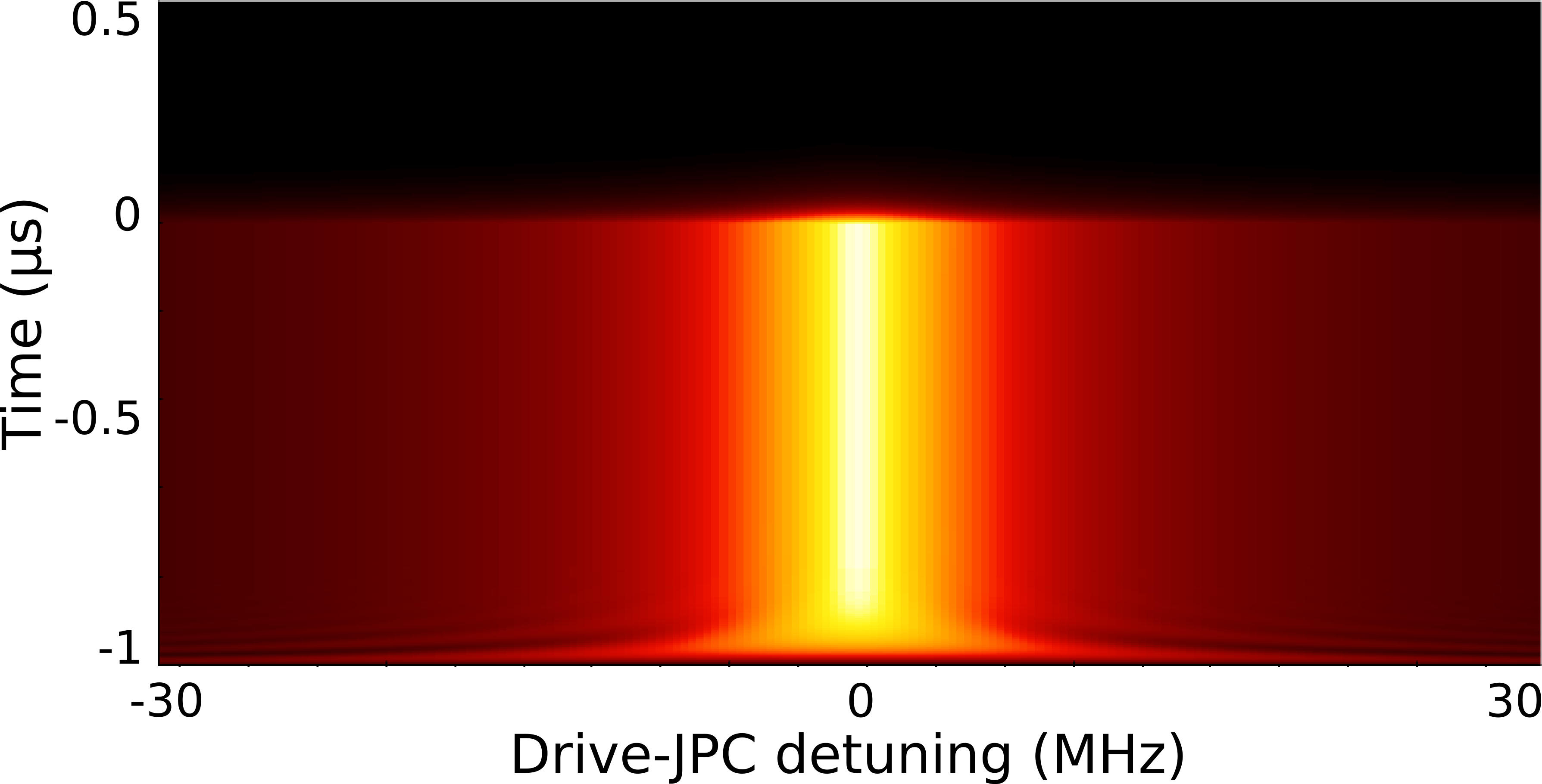}
\caption{Numerically simulated output field of the waveguide $\left<a_{out}\right>$ during a 1$\mu$s pulse: \textbf{a} the drive is resonant with the qubit, while sweeping the qubit-JPC. \textbf{b}, The same but without the qubit. In both pictures the pulse starts at t\,=\,-1$\mu$s and stops at 0.}
\label{theory_full_pulse}
\end{figure*}

In order to simulate the time-dynamics during the onset of the drive pulse, the system was also modeled numerically using QuTiP \cite{qutip1,qutip2}. The model, accounting for multiple qubits, is a modification of the model described in \cite{lalumiere_input-output_2013} to add the direct 'dipole-dipole' coupling term between the qubits $\mathcal{G}_{ij}$:

\begin{equation}
H =H_{qubits}+H_{qubit-drive}+H_{qubit-qubit}
\end{equation}
\begin{equation}
H_{qubits} = \sum_{i=1}^2\omega_{i} \left|e_i\right> \left<e_i\right|
\end{equation}
\begin{equation}
H_{qubit-drive} =  \sum_{i=1}^2\left(\epsilon_i \sigma_i^{+}+h.c.\right)
\end{equation}
\begin{equation}
H_{\textnormal{qubit-qubit}} = \sum_{j> i=1}^2\left( J_{ij} + \mathcal{G}_{ij}\right) \left( \sigma_i^- \sigma_j^+ + \sigma_i^+\sigma_j^-\right)
\end{equation}
\begin{equation}
\dot\rho = -\frac{i}{\hbar}\left[H,\rho\right] + \sum_{i,j=1}^2 \gamma_{ij}\left[ \sigma_i^- \rho \sigma_j^+ -\frac{1}{2}\lbrace \sigma_i^+\sigma_j^-, \rho\rbrace \right].
\end{equation}
\noindent Here $\omega_{i}$ is the frequency of the i-th qubit, $g_{i}$ is its dimensionless coupling to the waveguide, $\epsilon_i$ is the drive amplitude at the position of the i-th qubit, $J_{ij}$ is the exchange interaction amplitude and $\gamma_{ij}$ are the elements of the radiative decay matrix. In the case when two qubits are tuned in resonance and only driven from one side, the expressions for $J_{ij}$, $\gamma_{ij}$ and $\epsilon_{ij}$ are:
\begin{equation}
\epsilon_i = -i\sqrt{ \frac{\gamma^0_{ii}\omega}{2\omega_i}}\left<a_{in}\right>\textnormal{e}^{-i \omega t_i}
\end{equation}
\begin{equation}
J_{i j} = 2\pi g_i g_j \omega_i \textnormal{sin}\left( \omega_i t_{i j}\right)
\end{equation}
\begin{equation}
\gamma_{i j} =  4\pi g_i g_j \omega_i \textnormal{cos}\left( \omega_i t_{i j}\right) + \delta_{ij}\gamma_{internal} 
\label{eqn:gamma_ii}.
\end{equation}
Here $t_i=x_i/\nu$ defines the phase of the input signal $\left<a_{in}\right>$ propagating through the waveguide with a speed $\nu$ seen by i-th qubit located at point $x_i$ along the waveguide. Similarly, $t_{ij}=|x_i-x_j|/\nu$ defines the time delay of a signal propagating from the i-th to the j-th qubit. As before, $\omega$ is the drive frequency and $\gamma_{internal}$ is the non-radiative decay rate. In turn, $\gamma_{ii}^0$ is the radiative decay rate of the i-th qubit, i.e. $\gamma_{ii}$ in equation~\ref{eqn:gamma_ii} without $\gamma_{internal}$. In our configuration the qubits are located in a 3D waveguide effectively forming a closely packed system. In this case, the coupling term obtained form the waveguide input-output formalism gives $J_{ij}$=0 since the near-field components of the electromagnetic field are not considered. Such near-field term is instead included by introducing a direct qubit-qubit coupling term, $\mathcal{G}_{ij}$, which is related to the real part of the free space Green function. The output field is defined in the usual way, which is: 
\begin{equation}
\left<a_{out}\right> = \left<a_{in}\right> + \sum_{i=1}^2 \textnormal{e}^{i \omega_i t_i} \sqrt{\frac{\gamma^0_{ii}}{2}}\left<\sigma_i^-\right>.
\label{eqn:output}
\end{equation}
Solving these equations numerically allows us to reconstruct the field propagating along the waveguide regardless of the length of the drive pulse. This method was used for state the tomography presented in figure 3(a,b) of the main text. Figure \ref{theory_full_pulse} shows $\left<a_{out}\right>$ during the entire 1$\mu$s pulse: (a) with qubit, and (b) without qubit.
\section{Two-"atom" hybridization}
\label{app_F}
\begin{figure*}[t]
\raggedright
\textbf{a}\hspace{8.9cm}\textbf{b}\par
\includegraphics[width=0.5\linewidth]{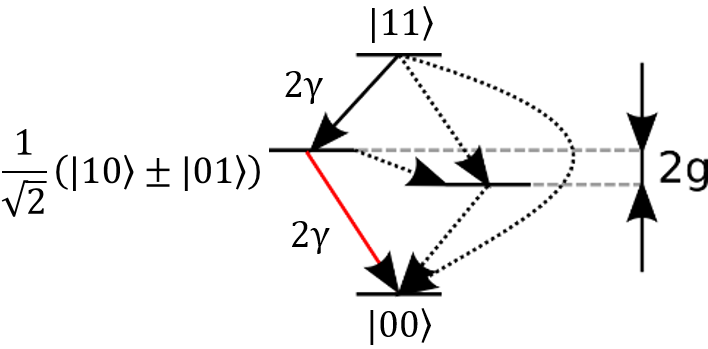}\hspace{1cm}
\includegraphics[width=0.4\linewidth]{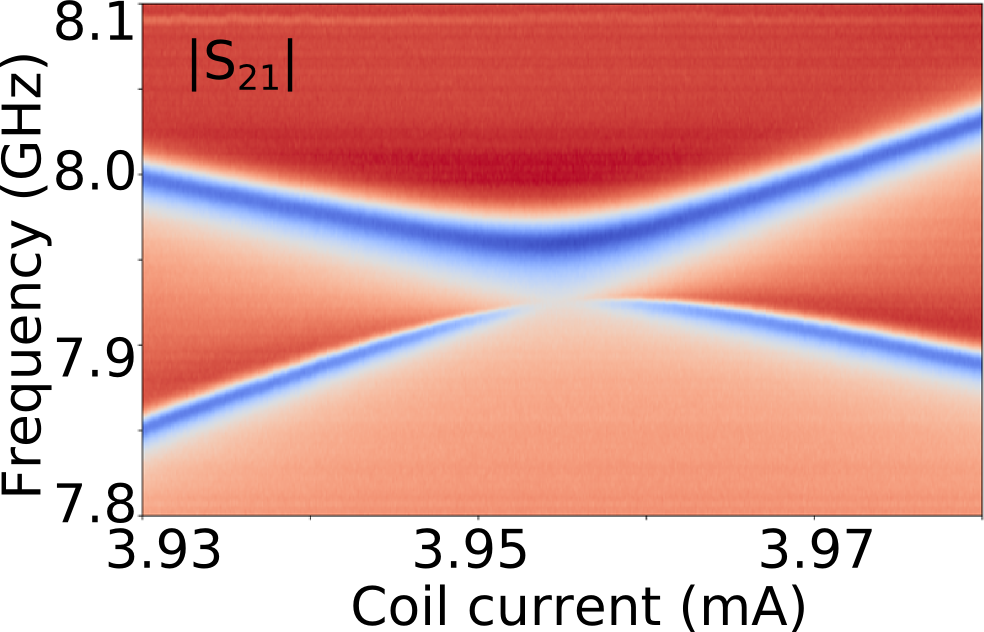}
\caption{\textbf{a}, Energy diagram of the first levels of two coupled transmons. Direct capacitive ('dipole-dipole') coupling provides levels hybridization and splitting $2g$. The red arrow corresponds to the transition we are using in the experiment. \textbf{b}, Experimentally measured transmission $|S_{21}|$ through the waveguide with 2 capacitively coupled transmons. Colour code from blue to red corresponds to $|S_{21}|$ change from 0 to 1. Both transmons are frequency tunable and coupled inductively to a magnetic coil, with a current flowing through it (swept along X-axis). The geometry of the system is different from the one used in the main text, to emphasise the dark and bright collective states of the system.}
\label{molecule}
\end{figure*}

Two resonantly interacting transmon qubits in the rotating wave approximation can be described by the interaction Hamiltonian $H_{int}=g(\sigma^{+}_1\sigma^{-}_{2}+h.c.)$, which in the single excitation manifold has two eigenstates $\frac{1}{\sqrt{2}}(\ket{01}\pm\ket{10})$. When two identical transmons are placed symmetrically in the waveguide, within the electric dipole approximation it is easy to show, that the dashed transitions in figure \ref{molecule}(a) are prohibited. Indeed, the transition probability is proportional to the square of the corresponding matrix element of the electric dipole operator $\sim |d_{fi}|^{2}$. This element can be expressed in terms of the wave functions of both transmons, given in charge representation. For example for the $\ket{11}\rightarrow \ket{01}\pm\ket{10}$ transition it reads:
\begin{equation}
    \begin{split}
        &d_{\ket{11}\rightarrow \ket{01}\pm\ket{10}}=\\
        &=\frac{l}{\sqrt{2}}\int \left(\psi_{\ket{1}}(Q_{1})\psi_{\ket{0}}(Q_{2})\pm\psi_{\ket{0}}(Q_{1})\psi_{\ket{1}}(Q_{2})\right) ^{*}(Q_{1}+Q_{2})\left(\psi_{\ket{1}}(Q_{1})\psi_{\ket{1}}(Q_{2})\right)dQ_{1}dQ_{2}=\\
        &\hspace{7cm}=\frac{1}{\sqrt{2}}(d_{\ket{1}\rightarrow\ket{0}}\pm d_{\ket{1}\rightarrow\ket{0}})=\sqrt{2}d_{\ket{1}\rightarrow\ket{0}}\;or\;0
    \end{split}
    \label{dipole_moment}
\end{equation}
where $l$ is proportional to the transmon length and $Q_{1}(Q_{2})$ is the charge operator of the first (second) transmon. According to equation \ref{dipole_moment} the probability of allowed transition is, in turn, increased twofold with respect to the single transmon case. States $\frac{1}{\sqrt{2}}(\ket{01}\pm\ket{10})$ are therefore referred as \textit{dark} and \textit{bright} respectively. Note, that their corresponding eigenvalues depend on the sign of the interaction constant $g$. The experiment with a photon emitted from a collective 2-transmon state has been done using the transition marked in figure \ref{molecule}(a) with a red arrow. Figure \ref{molecule}(b) shows transmission measurements through the waveguide with 2 transmons. The color code on the picture corresponds to the transmission coefficient $|S_{21}|$. The current in a coil coupled to both qubits is swept along the X-axis. To emphasise the difference between dark and bright states around the avoided crossing for this measurements, the qubits were installed in a way to have a larger coupling to the waveguide.


\end{document}